\documentclass{aa}  

\usepackage{graphicx}
\usepackage{txfonts}
\usepackage{xcolor}
\definecolor{xlinkcolor}{cmyk}{1,1,0,0}

 \usepackage[
   colorlinks=true,    
  linkcolor=xlinkcolor,     
 citecolor=xlinkcolor,     
filecolor=xlinkcolor,  
urlcolor=xlinkcolor
]{hyperref}
\usepackage{mathabx}
\begin{document}

   \title{The influence of a static planetary atmosphere on spin transfer during pebble accretion}

   \author{M.J. Yzer
          \and
          R.G. Visser
          \and
          C. Dominik
          }

   \institute{Astronomical Institute Anton Pannekoek, University of Amsterdam,
              Science Park 904, PO box 94249, Amsterdam, The Netherlands\\
              \email{[mitchell.yzer@student.;r.g.visser@;c.dominik@]uva.nl}
             }

   \date{}

 
  \abstract
   {Pebble accretion has been used to explain the small size of Mars, the heavy element contents of the gas giants, and the size distribution of asteroids. More recently, pebble accretion has been proposed as a means to explain not only the growth but also the prograde spin preference of most larger bodies in the Solar System. Pebble accretion could induce planetary and asteroid spin equal to or exceeding the spins currently measured. However, as these planetesimals grow, they start condensing the gas of the disc around them, forming an atmosphere within their Bondi radius.}
   {We study the effect an atmosphere has on the pebble orbits and spin build-up on the planet's surface during pebble accretion in the extreme case of a static atmosphere. Pebble feedback to the gas is not taken into account.}
   {The equations of motion for pebbles in a planar, global frame with a planet and a central star were integrated using the AIS15 integrator of REBOUND. An adiabatic atmosphere was then placed around the planet, and the spin deposited on the planet's surface was measured. These simulations were evaluated for different distances to the star, Stokes numbers, and planet masses.}
   {In general, an atmosphere dampens the spin the planet's surface receives by absorbing part of the angular momentum of the pebbles and circularising their orbits. This could prevent the excessive spin values predicted in some 3D pebble accretion simulations without an atmosphere. For planets larger than $0.5\ M_\Earth$, a stationary atmosphere absorbs all angular momentum, leaving no spin for the surface. Significant quantities of angular momentum are stored in the inner and intermediate atmosphere ($<0.3$ Bondi radii). Depending on the atmospheric and disc model, this spin could be transported either to the disc through atmospheric recycling or to the planet through drag between the surface and the atmosphere. Further research is required to quantify the spin transfer within the atmosphere.}
   {}

   \keywords{planets and satellites: formation --
                planets and satellites: atmosphere --
                methods: numerical --
                protoplanetary disk
               }

   \maketitle
%

\section{Introduction}
In the past decade, the theory of pebble accretion (PA) has gained significant traction. This theory proposes accretion regimes in which planetary seeds efficiently accrete aerodynamically small particles due to a combination of gravitational and dissipative forces. The process, therefore, requires both a planetary seed or planetesimal that is large enough to gravitationally attract the particles around it and pebbles that are small enough for their motion to be significantly altered by drag due to the gas in the disc \citep{Ormel_2017_PA}. In a gas-free scenario, accretion can only occur when a particle is on a trajectory that directly collides with the surface of the planetesimal. In the presence of gas, drag allows the pebble to settle in the gravitational field of the planetesimal. As it settles, the pebble slowly spirals inwards until it accretes onto the planetesimal. The accretion rate no longer depends on the planet's physical radius but on its mass \citep{Ormel_2017_PA}. In certain situations, the settling of pebbles can increase the accretion rate significantly compared to the gas-free scenario \citep{Ormel_2010}, since it can lead to a cross-section as large as the Hill sphere \citep{Lambrechts_2010}. Additionally, unlike planetesimals, pebbles are highly mobile in the disc and can drift from outer regions of the disc to the inner disc \citep{Weidenschilling_1977a}, replenishing the pebble reservoir and allowing for further growth.\\

Pebble accretion has become especially interesting with the discovery of large reservoirs of centimetre-size particles in protoplanetary discs \citep{Testi2003,Wilner_2005,Ricci_2010}. Since then, PA has been used to explain the heavy element contents of the gas giants \citep{Lambrechts_2014}, the small size of Mars \citep{Morbidelli_2015}, and the size distribution of the asteroids in the Solar System \citep{Johansen_2015b}.\\  

More recently, PA has been proposed as a means of spinning up the planets and large bodies in the Solar System. The bodies have a strong prograde rotation preference. Apart from Venus, all planets have their spin and orbital angular momentum axes aligned. The gas giants likely owe their spin to the angular momentum of gas that formed a circumplanetary disc and accreted onto the planet \citep{Machida_2008}. The story of the spin of the terrestrial planets and the ice giants is more complicated, partially since their spin might have been significantly influenced by post-formation processes \citep{DOBROVOLSKIS_1980,Laskar_1993}. The classical planet formation model, in which planetesimals grow due to collisions with other planetesimals \citep{Pollack_1996}, cannot explain planetary cores with strong systematic prograde spins \citep{Ida_1990,Lissauer_1991,Lissauer_1997,Dones_1993a}. This is mainly because prograde and retrograde spin contributions of different planetesimals nearly completely cancel each other out.\\

The current paradigm is therefore that the planets' spin is stochastically determined, resulting from large impacts at a late stage of planetary evolution \citep{Dones_1993b}. In these impacts, for example the hypothesised impact that formed the Moon, a Mars-sized embryo collides with the planet \citep{Wetherill_1958} at velocities close to the escape velocities at the surface of the planets, leaving the planet spinning at a frequency close to the break-up frequency, $\omega_\mathrm{crit}=\sqrt{GM_\mathrm{p}/R^3_\mathrm{p}}$ \citep{Kokubo_2007,Miguel_2010}, where $G$ is the gravitational constant and $M_\mathrm{p}$ and $R_\mathrm{p}$ are the mass and radius of the planet. The problem with this model is that it predicts an isotropic spin vector distribution, with the spin axis pointing towards low ecliptic latitudes, as Uranus's spin axis does. This is not the case for the other planets. Generally, this is not considered a problem because of the stochastic nature of the process, the small sample size of planets in the Solar System, and the aforementioned post-formation processes that alter the spin axis orientation.

Looking at the spin distribution of asteroids with diameters larger than 120 km, however, there is a strong non-uniformity in measured spins that cannot be explained by a collisional model \citep{Visser_2019}. These large asteroids are expected to have largely retained their primordial spin state \citep{Bottke_2005, Steinberg_2015}, and these states point to a systematic and inherent preference of the objects in the Solar System towards prograde spin.\\

\citet{Visser_2019} therefore propose that asymmetries in PA could result in an imbalance in prograde and retrograde spin contributions, leading to a significant net spin. This idea was first studied by \citet{Johansen_Lacerda_2010}, who used hydrodynamical simulations of accreting particles to determine the angular momentum transfer during PA. Using a high density of particles and a two-way drag force coupling between the pebbles and the gas, \citet{Johansen_Lacerda_2010} found that particle streams spiralled into a prograde circumplanetary disc, explaining the prograde rotation of the largest bodies in the asteroid and trans-Neptunian belts. These results were very insightful and promising, but they relied on very specific conditions. 

\citet{Visser_2019} therefore took a different approach. They used numerical simulations, integrating the equations of motion of individual pebbles and measuring the angular momentum the pebbles deposit on the surface, for both a 2D and 3D pebble distribution. Though these `less sophisticated' simulations did not capture the full dynamics between the gas and the pebbles as accurately as the hydrodynamical simulations of \citet{Johansen_Lacerda_2010}, ignoring among other things the back-reaction of the pebbles on the gas, they did allow for the study of a broad parameter space. \citet{Visser_2019} find that, in certain regions of the parameter space, there is significant angular momentum build-up on the surface, leading to prograde spins of the order of the current spins of the planets. With these results, \citet{Visser_2019} show that PA is not solely a viable mechanism to grow planets, but a means to provide the planets with their spin as well.\\

However, the combination of growth and spin transfer brings with it another problem. As the planets grow more massive, they start to condense the gas of the disc around them, forming a primordial atmosphere. This atmosphere in turn alters the drag that pebbles experience, which influences the amount of angular momentum the pebbles retain.\\

In a recent study, \citet{Takaoka_2023} used a combination of hydrodynamical and N-body simulations to show that, for a planet with a strongly prograde rotating, isolated atmosphere, pebbles provide the planet with a strong prograde spin, irrespective of the planet mass. Since the envelope becomes thicker and its prograde rotation becomes larger as the planet grows more massive, the authors identify a critical planet mass, after which the induced planetary spin by PA exceeds the break-up spin of the body.\\   

In this paper we focus on the other extreme: a fully static adiabatic atmosphere. We study the effect this atmosphere has on the pebbles' orbit and the angular momentum they transfer to the planet's surface. We use numerical simulations, similar to those of \citet{Visser_2019}, integrating the equations of motion of individual pebbles to determine the spin they deposit on the surface. By integrating over the full collision cross-section, we find the net spin in the presence and absence of an atmosphere. We perform a parameter study for different planet masses, distances to the central star, and Stokes numbers to identify large-scale trends. We show that, contrary to the rotating atmospheres from \citet{Takaoka_2023}, a static atmosphere can absorb significant fractions of the pebbles' spin, largely by circularising the pebbles' orbit. Though our simulations ignore feedback effects and angular momentum transfer within the atmosphere after the spin is deposited by the pebbles, we show where in the atmosphere the angular momentum is lost and discuss further evolution.

The structure of the paper is as follows. Section \ref{sec:numericalmodel} presents the numerical model, including the disc model (Sect. \ref{sec:discmodel}), atmospheric model (Sect. \ref{sec:atmosphericmodel}), initial conditions, and assumptions. In Sect. \ref{sec:results} the results are shown, starting with a fiducial model (Sect. \ref{sec:fiducialmodel}) to discuss the important mechanics within a simulation, followed by parameter studies for different distances to the star and planet masses (Sect .\ref{sec:spin_afo_a}) as well as for different Stokes numbers (Sect. \ref{sec:spin_afo_M_and_Stk}). These results are discussed in Sect. \ref{sec:discussion}. Finally, the most important conclusions are presented in Sect. \ref{sec:conclusion}.

\section{Model setup and initial conditions}
\label{sec:numericalmodel}

%
\subsection{Disc model}
\label{sec:discmodel}
Following from the minimum mass solar nebula \citep{Weidenschilling_1977b, Hayashi, NAKAGAWA1986}, the temperature and surface density profiles of the protoplanetary disc are
\begin{align}
        T(r) &= 170 \mathrm{\ K} \left( \frac{r}{1 \mathrm{\ au}}\right)^{-1/2},\\
        \Sigma(r) &= 17000 \mathrm{\ kg\ m}^{-2} \left( \frac{r}{1 \mathrm{\ au}}\right)^{-3/2},
\end{align}
with $r$ the distance to the star. Assuming hydrostatic equilibrium and an isothermal temperature profile in the vertical disc direction, the density profile is given by
\begin{equation}
        \rho_\mathrm{g}(r,z)=\frac{\Sigma(r)}{H\sqrt{2\pi}}\exp \left\{-\frac{1}{2}\left(\frac{z}{H}\right)^2\right\},
\end{equation}
in which $z$ is the vertical position compared to the mid-plane and the scale height $H=c_\mathrm{s}/\Omega_0$. In this expression $\Omega_0$ is the local Keplerian frequency and $c_\mathrm{s}$ is the local sound speed, given by $c_\mathrm{s}=\sqrt{k_\mathrm{b}T/\mu}$, with $k_\mathrm{b}$ the Boltzmann constant, $T$ the temperature and $\mu$ the mean molecular mass.

The gas of the disc is pressure supported. As a result of this, the orbital velocity of the gas is slightly less than Keplerian. Solid objects like planetesimals, therefore, experience a constant headwind of magnitude
\begin{equation}
        v_\mathrm{hw} = \frac{nk_\mathrm{b}T}{2\mu V_\mathrm{k}},
\end{equation}
with $V_\mathrm{k}$ being the velocity of a circular Keplerian orbit and $n\approx 3.25$ the exponent of the pressure power-law profile of the gas \citep{Weidenschilling_1977a}. Due to this headwind, pebbles experience drag, as a result of which they lose angular momentum and spiral inwards.

\subsection{Atmospheric model}
\label{sec:atmosphericmodel}
The addition of a static planetary atmosphere increases the gas density in the vicinity of the planet, increasing drag as the pebbles make a close approach. The atmosphere in this study consisted of a single, adiabatic layer.\\

A planet can condense and bind the gas within its Bondi radius, given by
\begin{equation}
        \label{eq:Rbondi}
        R_\mathrm{B}=\frac{G\mu M_\mathrm{p}}{k_\mathrm{b}T_\mathrm{d}},
\end{equation}
with $G$ the gravitational constant, $\mu$ the mean molecular mass of the gas, $M$ the mass of the planet and $T_\mathrm{d}$ the temperature of the disc at the location of the planet.   
The outer edge of the atmosphere $R_\mathrm{out}$ is determined by the lesser of the Bondi radius ($R_\mathrm{B}$) and the Hill radius ($R_\mathrm{H}$) \citep{Pollack_1996, Benvenuto2005, HUBICKYJ2005415, Lee_2015}. The Hill radius is given by 
\begin{equation}
        \label{eq:Rhill}
        R_\mathrm{H} = a_\mathrm{p}\sqrt[\leftroot{-1}\uproot{2}\scriptstyle 3]{\frac{M_\mathrm{p}}{3M_\mathrm{star}}},
\end{equation}
with $a_\mathrm{p}$, also referred to as a$_\mathrm{plan}$, the distance from the planet to the star and $M_\mathrm{star}$ the mass of the star.\\


In our model, we assumed a single-layered metal-free convective atmosphere. The temperature, density and pressure profiles of this atmosphere are given by \citep[see e.g.][]{Brouwers_2020}

\begin{subequations}
        \label{eq:atmosphericconditions}
        \begin{align}
                T_\mathrm{a}(r) &= T_\mathrm{d}\left(1+\frac{\gamma -1}{\gamma}R_\mathrm{B}\left( \frac{1}{r}- \frac{1}{R_\mathrm{out}}\right)\right),\\
                \rho_\mathrm{a}(r) &= \rho_\mathrm{d}\left(1+\frac{\gamma -1}{\gamma}R_\mathrm{B}\left( \frac{1}{r}- \frac{1}{R_\mathrm{out}}\right)\right)^{\frac{1}{\gamma - 1}}, \\
                P_\mathrm{a}(r) &= P_\mathrm{d}\left(1+\frac{\gamma -1}{\gamma}R_\mathrm{B}\left( \frac{1}{r}- \frac{1}{R_\mathrm{out}}\right)\right)^{\frac{\gamma}{\gamma - 1}},
        \end{align}
\end{subequations}
in which $T_\mathrm{a}$, $\rho_\mathrm{a}$ and $P_\mathrm{a}$ refer to the temperature, density and pressure within the atmosphere and $T_\mathrm{d}$, $\rho_\mathrm{d}$ and $P_\mathrm{d}$ refer to the conditions in the disc. $r$ is the distance to the planet's surface, $R_\mathrm{out} = \min(R_\mathrm{B}, R_\mathrm{H})$ and $\gamma = 1.45$ is the adiabatic index of the gas of the disc. 

\begin{figure*}
        \centering
        \includegraphics[width=\linewidth]{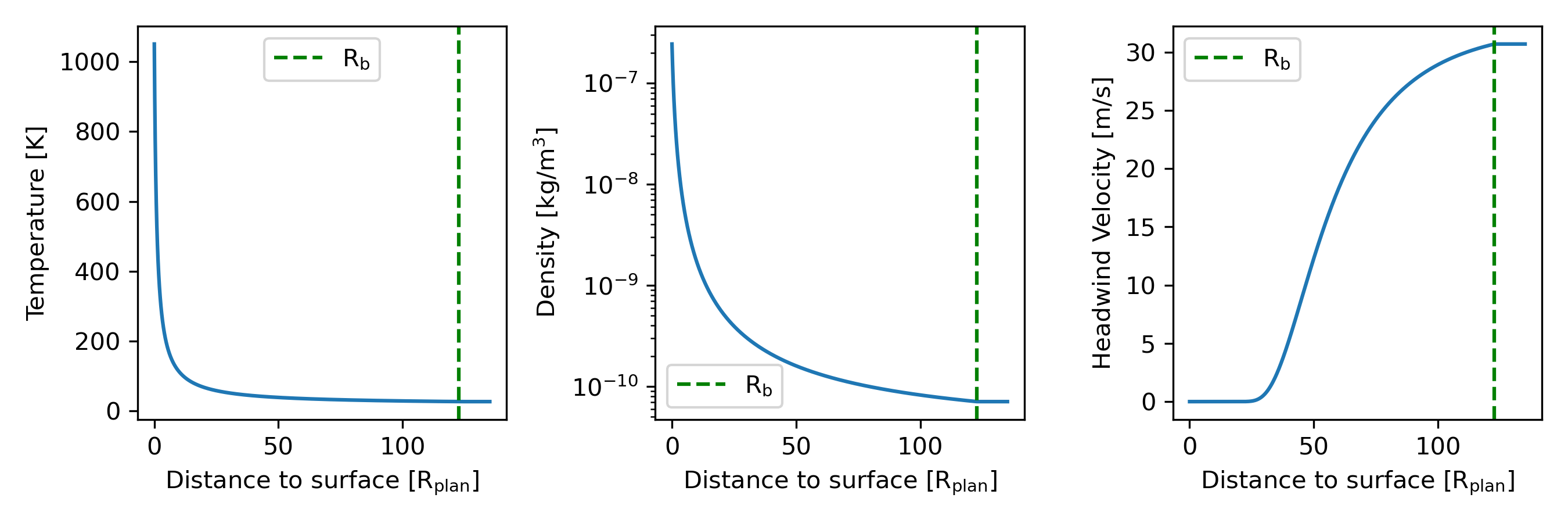}
        \caption{Atmospheric and disc conditions in the fiducial model ($a_\mathrm{p}=40$ au, $M_\mathrm{p}=0.1 M_\Earth$, $\tau_\mathbf{s}$ = 0.1) as a function of distance to the planet's surface. The distance to the surface is expressed in terms of the planet's radius. The dashed green line indicates the Bondi radius, which is the outer edge of the atmosphere. The expressions for these curves are given in Eqs. \ref{eq:atmosphericconditions} and \ref{eq:fullheadwind}.}
        \label{fig:atm_cond}
\end{figure*}

Since the atmosphere is stationary around the planet, the tenuous headwind disappears once the pebble is deep within the atmosphere. Previous studies like \citet{okamura2021} have neglected the headwind. However, according to \citet{Ormel_2014_atm_rec_I}, the headwind strongly affects the flow pattern in and around the atmosphere. We, therefore, did not ignore the headwind but used an inverse exponential to let it fall off. The headwind is then given by
\begin{equation}
        \label{eq:fullheadwind}
        v_\mathrm{hw} =  \frac{nk_\mathrm{b}T_\mathrm{d}}{2\mu V_\mathrm{k}} \exp \left\{-\frac{\rho_\mathrm{a}(r)-\rho_\mathrm{d}}{\rho_\mathrm{d}}\cdot \frac{0.3 R_\mathrm{B}}{r}, \right\}
,\end{equation}
in which $\rho_\mathrm{a}=\rho_\mathrm{d}$ for $r\ge R_\mathrm{out}$ and in which $0.3 R_\mathrm{B}/r$ is a factor that modulates the point at which the headwind approaches zero. This point is believed to be around 0.3 $R_\mathrm{B}$ \citep{Ormel_2015b}. An example of the temperature, density and headwind profiles for a typical atmosphere is shown in Fig. \ref{fig:atm_cond}.\\  

As mentioned in the introduction, in contrast to \citet{Takaoka_2023}, who studied the effects of a strongly rotating atmosphere, we focus on the extreme case of a static atmosphere. This reduces the asymmetry between prograde and retrograde orbiting pebbles and their interaction with the atmosphere. It is therefore to be expected that, whereas \citet{Takaoka_2023} report very strong prograde spin build-up, this study will show smaller spin values.

For simplicity, we assumed that the core is non-rotating as well. In reality, the spin of the initial core depends on its path of formation. It could be strongly rotating in the prograde direction if, for example, the core originated from the collapse of a pebble cloud \citep{Visser_2022} or it could be virtually standing still if planetesimal accretion was the main formation mechanism \citep{Ida_1990,Dones_1993a}. However, this assumption does not influence the dynamics of the simulations or the interpretation of the results, since we are focussing on the amount of angular momentum that is brought to the core by PA, which is independent of any spin a planet might already have. 


A more impactful assumption used in this study is that the atmosphere is fully convective, rather than consisting of an adiabatic inner and an isothermal outer layer, such as in the models of \citet{Brouwers_2020}. The fully convective atmosphere is valid for small planets ($\lesssim 2 {\rm\ M}_\Earth$) and large accretion rates ($\gtrsim 10^{-6}$ M$_\Earth$/yr), due to pebble and dust induced opacity enhancements \citep{Brouwers_2021}. Since typical accretion rates lie around $10^{-5}$ M$_\Earth$/yr \citep{Lambrechts_2014}, the adiabatic atmosphere should be valid for all planets used in this study. 

Finally, processes like the back-reaction of the pebbles on the gas, the flow of the disc gas around the atmosphere and the ablation of the pebbles were ignored.

\subsection{The equations of motion}
\label{sec:EoM}
To study the behaviour of pebbles that enter the atmosphere and determine the amount of spin they deposit on the planet's surface, we integrated the equations of motion in a global, planar (2D) model. The equations of motion of a pebble in this heliocentric frame are given by
\begin{equation}
        \begin{split}
                \frac{d^2\mathbf{r}}{dt^2} &= \mathbf{f}_\mathrm{star} + \mathbf{f}_\mathrm{planet} + \mathbf{f}_\mathrm{drag} \\
                &= -GM_\mathrm{s}\frac{\mathbf{r}}{r^3} - GM_\mathrm{p}\left(\frac{\mathbf{r}-\mathbf{r}_\mathrm{p}}{|\mathbf{r}-\mathbf{r}_\mathrm{p}|^3}+\frac{\mathbf{r}_\mathrm{p}}{r_\mathrm{p}^3} \right) -\frac{\mathbf{v}_\mathrm-\mathbf{v}_\mathrm{gas}}{t_\mathrm{s}}.
        \end{split}
\end{equation}
In this equation, $G$ is the gravitational constant, $M_\mathrm{s}$ and $M_\mathrm{p}$ are the masses of the star and planet, respectively, $\mathbf{r}$ and $\mathbf{r}_\mathrm{p}$ are the position vectors of the pebble and the planet, both with respect to the star, $\mathbf{v}$ is the velocity of the pebble and $t_\mathrm{s}$ is the stopping time of the pebbles. $\mathbf{v}_\mathrm{gas}$ is the velocity of the gas, given by
\begin{equation}
        \mathbf{v}_\mathrm{gas}=\left(V_\mathrm{k}-v_\mathrm{hw}\right)\textbf{e}_\phi,
\end{equation}
in which $\textbf{e}_\phi$ is the unit vector in the azimuthal direction.

Meanwhile, the stopping time $t_\mathrm{s}$ is dependent on the size of the pebbles and on the local disc conditions. Depending on the ratio between the particle size and the mean free path in the disc, the pebble drag is described by either the Epstein drag law, for aerodynamically small particles, or the Stokes drag law, for large particles. The Stokes drag regime is itself split up into three different regimes, depending on the Reynolds number, given by

\begin{equation}
        \mathrm{Re} = 2s\rho v_\mathrm{pg}/\eta,
\end{equation}
in which $s$ is the particle's radius, $\rho$ is the local density of the gas, $v_\mathrm{pg}$ is the speed of the pebble with respect to the local gas and $\eta$ is the gas viscosity.\\   

The stopping time is then given by \citep{Weidenschilling_1977a}
\begin{equation}
        t_\mathrm{s}=
        \begin{cases}
                \vspace{1.5 mm}\hspace{6.0 mm}\cfrac{\rho_\mathrm{s}s}{\rho \Bar{v}} \quad &\mathrm{for} \quad s < \frac{9}{4}\lambda\\
                \vspace{1.5 mm}\hspace{4.5 mm}\cfrac{2\rho_\mathrm{s}s^2}{9\eta} \quad & \mathrm{for} \quad s \ge \frac{9}{4}\lambda, \quad \mathrm{Re} \le 1\\
                \vspace{1.5 mm}\cfrac{2^{0.6}\rho_\mathrm{s}s^{1.6}}{9\eta^{0.6}\rho^{0.4}v_\mathrm{pg}^{0.4}} \quad & \mathrm{for} \quad s \ge \frac{9}{4}\lambda, \quad 1 < \mathrm{Re} < 800\\
                \hspace{5 mm}\cfrac{6\rho_\mathrm{s}s}{\rho v_\mathrm{pg}} \quad & \mathrm{for} \quad s \ge \frac{9}{4}\lambda, \quad \mathrm{Re} \ge 800.\\
        \end{cases}
\end{equation}
In these equations, $\rho_\mathrm{s}$ is the density of the pebbles, taken to be 3 g cm$^{-3}$, $\Bar{v}$ is the mean thermal velocity of the gas and $\lambda$ is the mean free path.

\begin{figure}
        \centering
        \includegraphics[width=\linewidth]{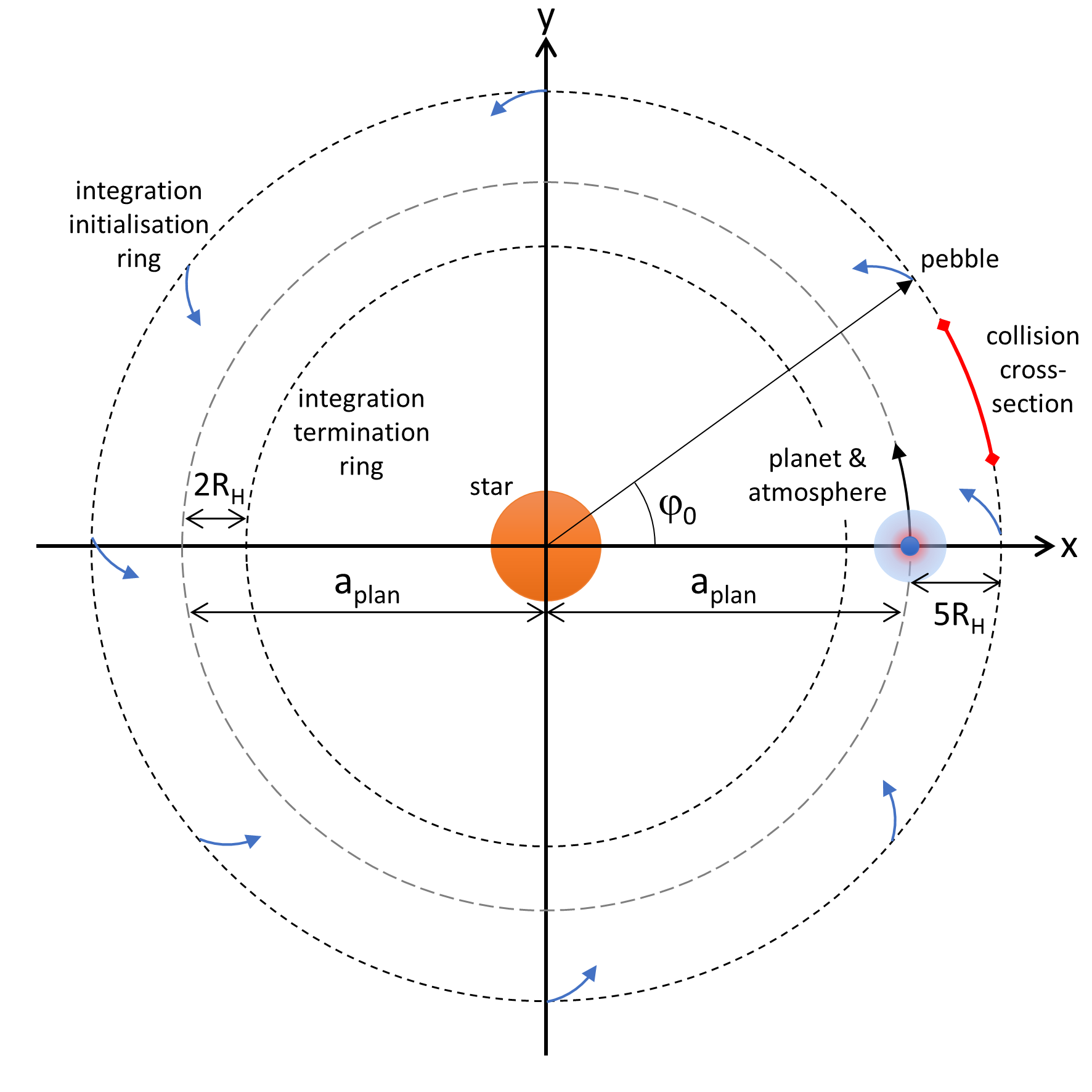}
        \caption{Schematic overview of the simulation setup. The planet is initialised on the x-axis on a circular orbit with radius a$_\mathrm{plan}$. The pebbles are released from an exterior ring with a radius of a$_\mathrm{plan}$ + 5$R_\mathrm{H}$ and slowly drift inwards. The pebbles are identified by their initial azimuthal coordinate, $\phi_0$. At the start of the simulation, the collision cross-section is calculated. Pebbles released from within this region, symbolised by the red ring segment, eventually accrete onto the planet. The integration of pebbles that miss the planet is terminated once they drift past the interior ring with a radius of a$_\mathrm{plan}$ - 2$R_\mathrm{H}$.}
        \label{fig:simsetup}
\end{figure}

At the start of an integration, the pebbles were initialised along a circular orbit exterior to the planet's orbit, with a radius of a$_\mathrm{plan}$ + 5$R_\mathrm{H}$. The pebbles were identified using their initial azimuthal coordinate $\phi_0$ along this arc. The integration of a pebble's orbit was terminated at the moment the pebble collided with the planet, given by $|\mathbf{r}_\mathrm{p}-\mathbf{r}|=R_\mathrm{p}$, or when it had drifted past the planet to an interior orbit with radius $r_\mathrm{terminate}=\mathrm{a}_\mathrm{plan}-2R_\mathrm{H}$. Figure \ref{fig:simsetup} provides a schematic representation of the simulation setup. The details of the numerical integration and initial conditions, as well as the caveats of this approach, are discussed in Appendix \ref{sec:appendix_A}.

\subsection{Spin transfer through pebble accretion}
\label{sec:model setup - spin transfer}
The spin transfer from a pebble to the planet was determined by calculating the specific angular momentum (SAM) of the pebble around the planet's centre at the moment of impact. The SAM that is transferred during this impact is given by
\begin{equation}
        \mathbf{l} = \mathbf{r} \times \mathbf{v},
\end{equation}
in which $\mathbf{r}$ and $\mathbf{v}$ refer to the position and velocity of the pebble with respect to the planet at the moment of impact. In this study, we focussed on spin transfer from pebbles in the ecliptic plane. Therefore, the only relevant component is the spin in the z-direction, given by
\begin{equation}
        l_\mathrm{z}= \Delta x \Delta v_\mathrm{y} - \Delta y \Delta v_\mathrm{x},
\end{equation}
in which $\Delta i = i_\mathrm{pebb} - i_\mathrm{plan}$ with $i\in\{x,y,v_\mathrm{x},v_\mathrm{y}\}$. By averaging these SAMs over the full collision cross-section \citep{Visser_2016}, we find the 2D mean SAM \citep{Visser_2019}:
\begin{equation}
        \langle l_\mathrm{z} \rangle = \cfrac{\int_\mathrm{col.\ c.s.}F(\phi_0)l_\mathrm{z}(\phi_0)\mathrm{d}\phi_0}{\int_\mathrm{col.\ c.s.}F(\phi_0)\mathrm{d}\phi_0},
\end{equation}
in which $F(\phi_0)$ is the flux of pebbles with an initial azimuthal coordinate $\phi_0$ within the collision cross-section and $l_\mathrm{z}(\phi_0)$ is the spin those pebbles provide to the planet. Because of the symmetry of the heliocentric system, $F$ is constant for all values of $\phi_0$ and therefore does not influence the mean SAM. To estimate this mean spin, 500 pebbles, linearly spaced throughout the collision cross-section, were released and their SAMs were averaged. The SAM is often expressed as a fraction of the escape SAM: $l_\mathrm{z,esc}=\sqrt{2GM_\mathrm{p}R_\mathrm{p}}$. This relates to the spin angular frequency through \citep{Visser_2019}
\begin{equation}
        \frac{\omega_\mathrm{z}}{\omega_\mathrm{crit}}\approx 3.5 \frac{\langle l_\mathrm{z} \rangle}{l_\mathrm{z,esc}},
\end{equation}
in which $\omega_\mathrm{z}$ is the spin angular frequency of the planet, provided there are enough pebbles in the disc to significantly spin up the planet, and $\omega_\mathrm{crit}=\sqrt{GM_\mathrm{p}/R_\mathrm{p}^3}$ is the break-up spin frequency.

\section{Results}
\label{sec:results}
This study aimed to analyse the influence of a static atmosphere on spin transfer during PA. To this end, the mean SAM transferred to the planet in the presence and absence of an atmosphere was measured for different orbital radii and masses of the planet, as well as for different Stokes numbers of the pebbles. The results are discussed in Sects. \ref{sec:spin_afo_a} and \ref{sec:spin_afo_M_and_Stk}. We assumed densities of $\rho$ = 3 g cm$^{-3}$ for both the pebbles and the planetesimals \citep{Liu&Ormel2018} and a stellar mass of 1 M$_\odot$. In the results of these parameter studies, the data from individual simulations are reduced to a single parameter, namely the mean SAM of the pebbles at the moment of impact with the planet, or equivalently the resulting spin angular frequency of the planet. In order to demonstrate the mechanics behind the amount of transferred spin, we first present a fiducial model. In this fiducial model, we discuss the effects of an atmosphere on a single simulation in more detail. In this particular simulation, a 0.1 Earth-mass planet, placed at 40 au from the star, is accreting pebbles with a Stokes number ($\tau_\mathrm{s}$) of 0.1. 

\subsection{Fiducial Model: a$_\mathrm{p}$ = 40 au, M$_\mathrm{p}$ = 0.1 M$_\Earth$, $\tau_\mathbf{s}$ = 0.1}
\label{sec:fiducialmodel}

\begin{figure*}[t]
        \centering
        \includegraphics[width=0.8\linewidth]{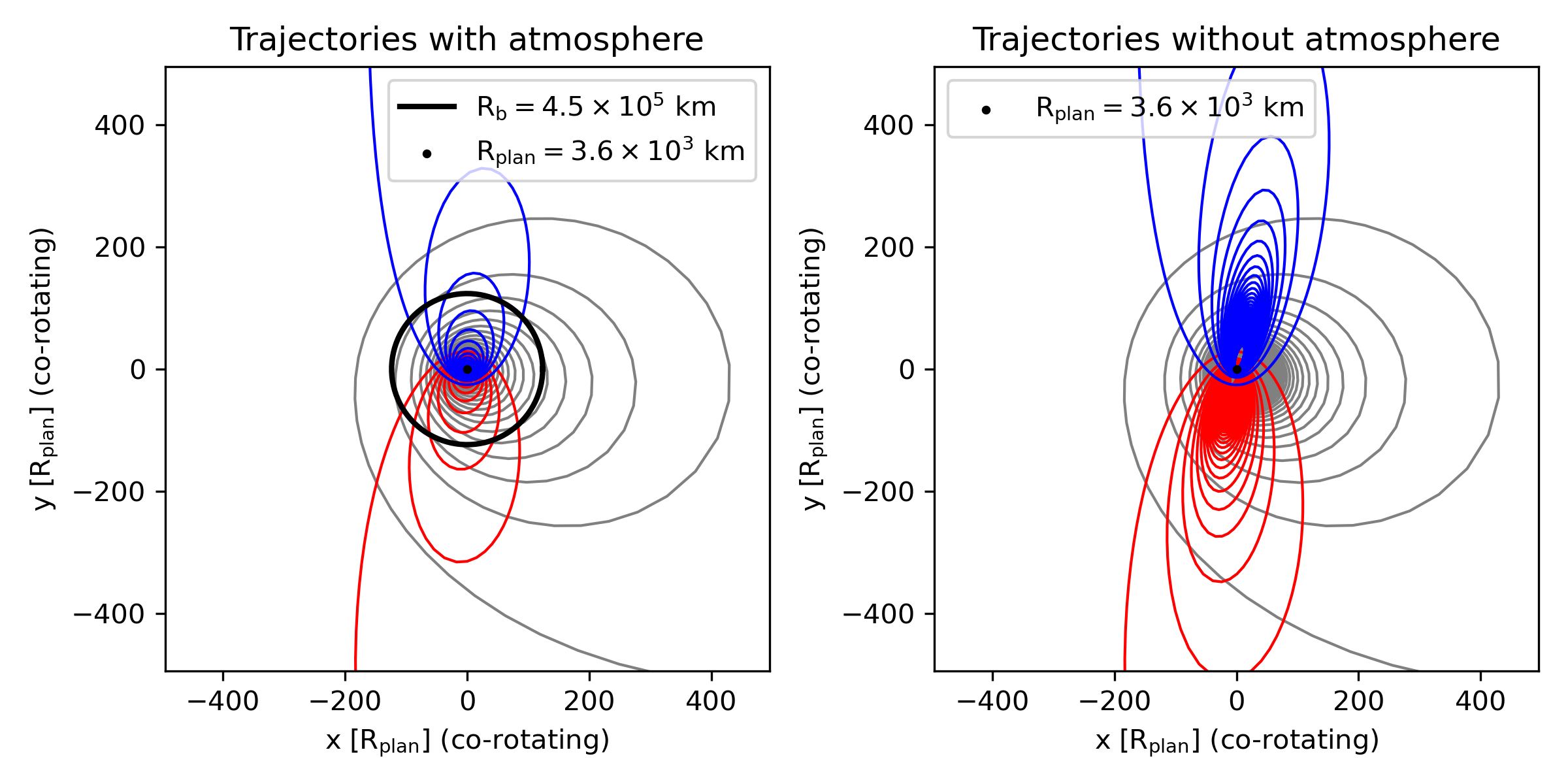}
        \caption{Trajectories of three pebbles in the fiducial model with and without an atmosphere. The trajectories show the final approach of the pebbles, deep within the Hill sphere. The coordinates are transformed from the global frame to a local frame co-rotating with the planet in order to highlight the motion of the pebbles with respect to the planet. The black circle in the left plot is the Bondi circle, indicating the start of the atmosphere. High-eccentricity orbits, shown with the red and blue curves, are circularised by the atmosphere and require significantly fewer orbits to accrete than their atmosphere-less counterparts, due to the high gas density at their pericentre. Low-eccentricity orbits, shown with the grey curve, are more subtly influenced, mainly in terms of the number of revolutions before impact.}     
        \label{fig:40AUfidmodeltraj}
\end{figure*}

\begin{figure*}
        \centering
        \includegraphics[width=0.8\linewidth]{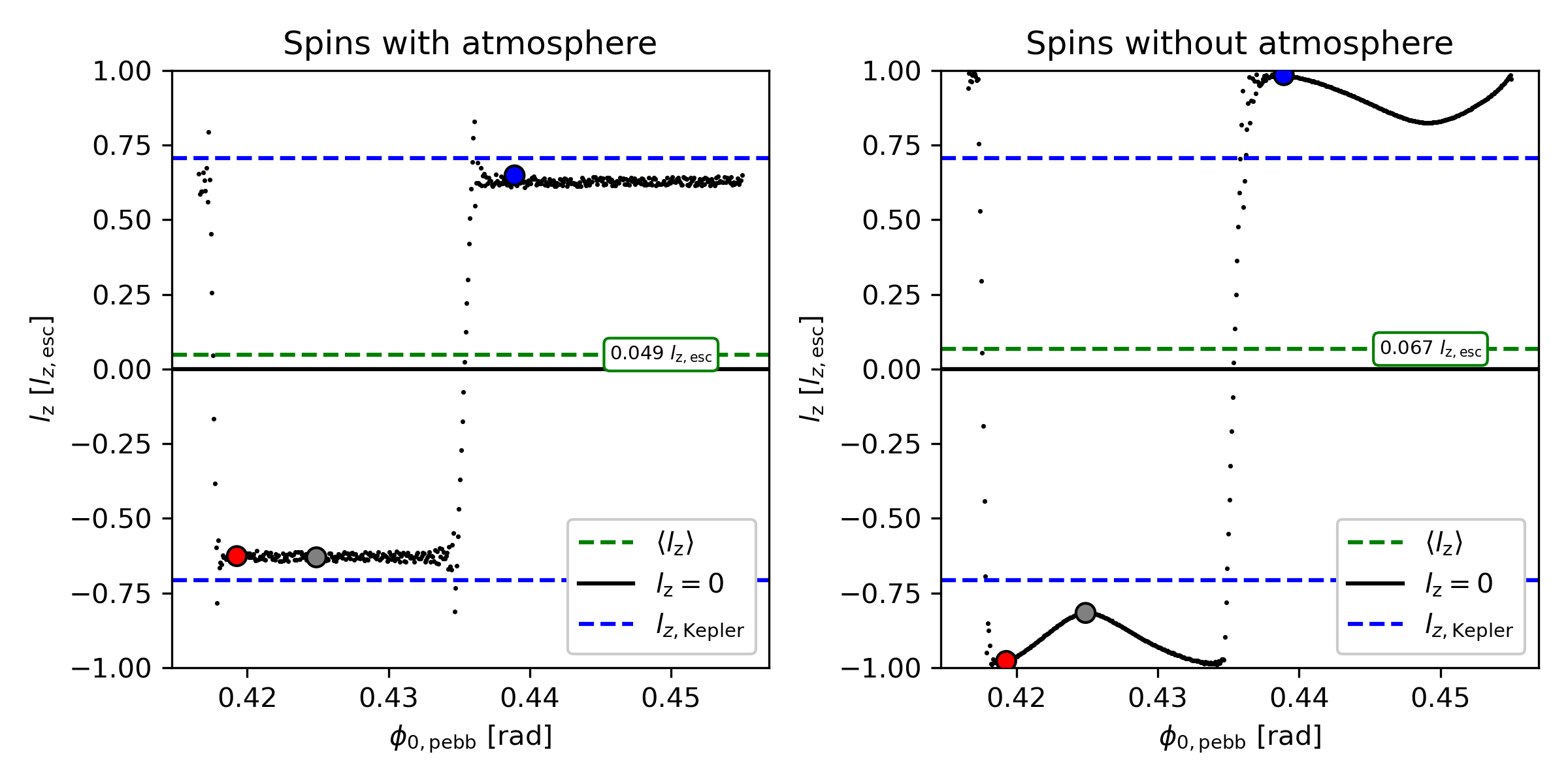}
        \caption{Angular momenta of the pebbles in the fiducial model at the moment of impact with the planet. The individual SAMs of the pebbles are plotted as a function of the pebbles' initial position within the collision cross-section. In these spin maps, the SAMs of the trajectories in Fig. \ref{fig:40AUfidmodeltraj} are highlighted with dots in the respective colours. A dashed green line shows the mean SAM, while the dashed blue lines show the angular momenta of a circular Keplerian orbit. The left panel shows the situation with an atmosphere, while the right panel shows the same pebbles without an atmosphere. The addition of an atmosphere reduces the mean angular momentum from $0.067$ to $0.049\ l_\mathrm{z,esc}$, corresponding to a spin frequency of the planet of 0.24 and 0.17 $\omega_\mathrm{crit}$, respectively. Without an atmosphere, the individual spin contributions are distributed between the escape angular momentum ($l_\mathrm{z}/l_\mathrm{z,esc}= \pm 1.00$) and the Kepler angular momentum ($l_\mathrm{z}/l_\mathrm{z,esc}= \pm \sqrt{2}/2$). In the presence of an atmosphere, nearly all spin contributions are reduced to a sub-Keplerian angular momentum plateau. This suggests a circularisation of the orbits, less tangential impacts, or a combination of the two.}     
        \label{fig:40AUfidmodelspin}
\end{figure*}

In the fiducial model, a 0.1 Earth-mass planet is following a circular orbit around a solar-mass star at a distance of 40 au. In the simulations, the planet is only accreting pebbles with a Stokes number ($\tau_\mathrm{s}$) of 0.1. These pebbles, 500 in total, are linearly spaced throughout the collision cross-section, as described in Sect. \ref{sec:model setup - spin transfer}. 
There is no mass transfer or planetary growth in the simulation. There is no accretion of gas either, only the condensation of the gas within the Bondi sphere in simulations in which the atmosphere is turned on.\\ 

The conditions of this particular fiducial model were chosen primarily for demonstrative purposes. There is a relatively strong variety in the shapes of the trajectories of the different pebbles throughout the collision cross-section in this model and the mean transferred spin is high. This means that the general mechanisms governing all simulations are most apparent in this model. On top of that, the conditions are similar to those of an ice giant far out in the disc, like Neptune, at a stage in which the planet is massive enough to create a significant atmosphere around it, but small enough to still be growing significantly. These gas giants are likely to develop a primordial atmosphere early on, meaning that the atmospheric influence on PA will be highly important, assuming PA plays a part in their growth.

\subsubsection{Pebble trajectories and spin contributions}

The main results of this model are seen in Figs. \ref{fig:40AUfidmodeltraj} and \ref{fig:40AUfidmodelspin}. Figure \ref{fig:40AUfidmodeltraj} shows the trajectories of three pebbles selected from within the collision cross-section with and without an atmosphere. These trajectories are transformed from the global frame to a local frame co-rotating with the planet in order to accentuate the motion of the pebbles relative to the planet. 

Figure \ref{fig:40AUfidmodelspin} shows the individual spin contributions of each of the 500 pebbles within the collision cross-section, with their initial azimuthal coordinate on the x-axis and their angular momentum with respect to the planet's centre at the moment of impact on the y-axis. Here, the angular momenta of a circular Keplerian orbit with a radius equal to the radius of the planet is given by $l_\mathrm{z,Kepler}=V_\mathrm{k}R_\mathrm{p}=\sqrt{GM_\mathrm{p}R_\mathrm{p}}$ (see the dashed blue lines). These types of spin maps are discussed in greater detail in \citet{Visser_2019} for different accretion regimes in systems without an atmosphere.\\

The most important piece of information is the mean angular momentum transferred by the pebbles. This is the main information used in Sects. \ref{sec:spin_afo_a} and \ref{sec:spin_afo_M_and_Stk}. The atmosphere reduces the mean spin from $0.067$ to $0.049\ l_\mathrm{z,esc}$, corresponding to a reduction in the resulting spin frequency of the planet from 0.24 to 0.17 $\omega_\mathrm{crit}$. This reduction in mean angular momentum comes from a reduction of the individual spin contributions of all pebbles, as can be seen when comparing the left and right panels of Fig. \ref{fig:40AUfidmodelspin}. The reduction is not the result of a significant re-balancing between the prograde and retrograde spin contributions.

In the system without an atmosphere, the SAMs are distributed between the escape angular momentum $(l_\mathrm{z}= \pm 1.00\ l_\mathrm{z,esc})$ and the Kepler angular momentum. In the presence of an atmosphere, however, most angular momenta are forced to one of two sub-Keplerian angular momentum plateaus. This suggests either a strong reduction in impact velocity or less tangential impacts. A strong impact velocity reduction could, in turn, suggest a circularisation of the orbits, since the argument of impact will be close to the argument of periapsis and a reduction in the velocity at periapsis can be ascribed to a reduction in eccentricity. However, there are other possible explanations for the velocity reduction, for example, extreme drag in the deepest parts of the atmosphere leading to a large speed reduction right before impact.\\

\begin{figure*}[t]
        \centering
        \includegraphics[width=0.95\linewidth]{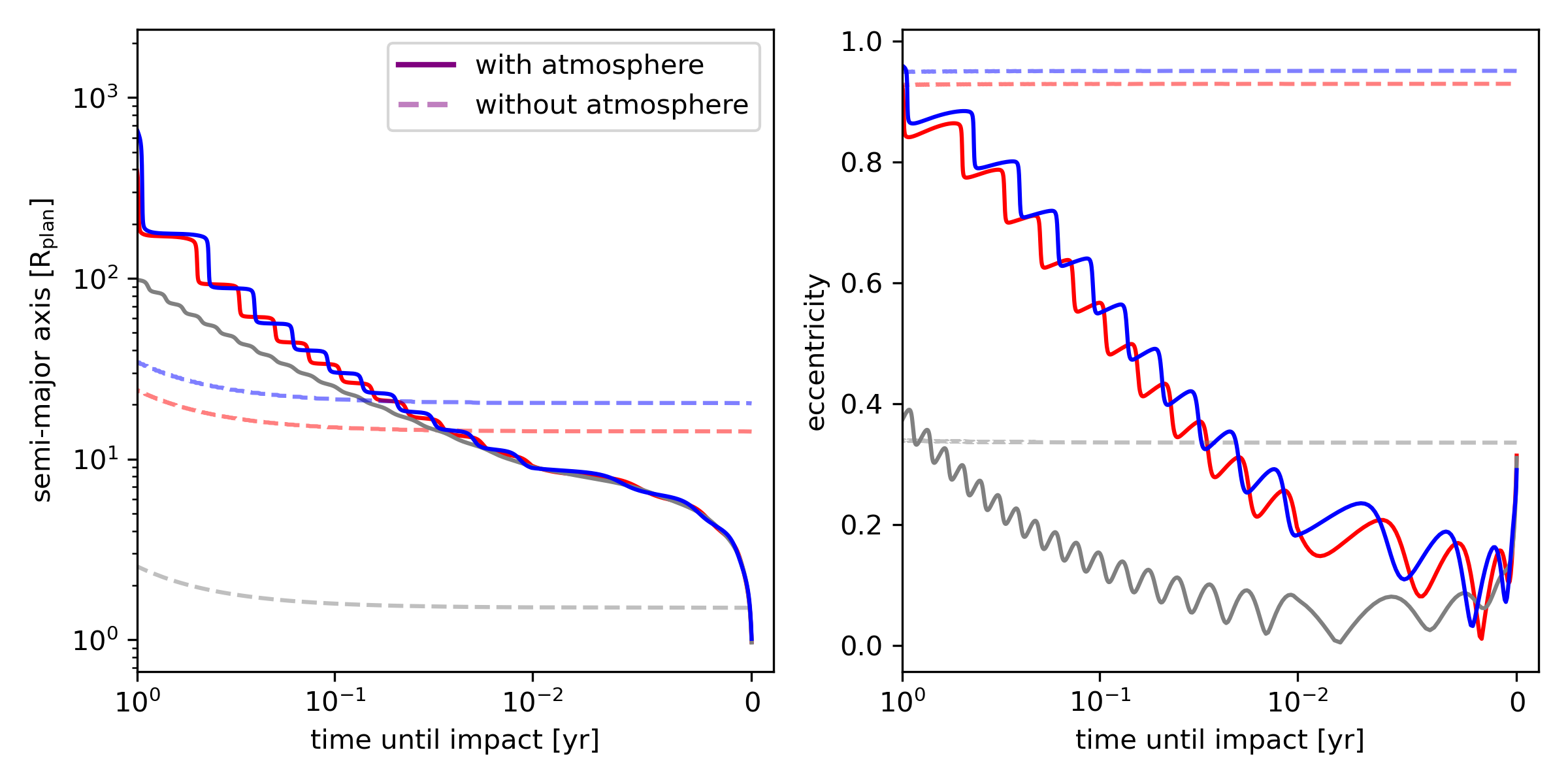}
        \caption{Evolution of the semi-major axis and eccentricity of the trajectories shown in Fig. \ref{fig:40AUfidmodeltraj}. The solid lines represent the pebbles with an atmosphere, while the faint, dashed lines represent the same pebbles without atmosphere. The x-axis shows the time until the impact and is scaled logarithmically between $10^0$ and $10^{-2}$ yr and linearly between $10^{-2}$ and 0 yr. With an atmosphere, the eccentricity is reduced to zero during the final stage before the impact, after which the eccentricity rapidly increases, indicating a phase of strong orbital decay. This rapid decay is seen in the semi-major axis as well. Without atmosphere, there is hardly any change in the orbital elements. Note that, though it might seem in the left plot as if the red and blue pebbles without an atmosphere do not come close enough to the planet to impact it, the orbits of these pebbles are so eccentric that even though their semi-major axes remain large, their pericentres exactly coincide with the planet's surface at $t=0$, leading to an almost perfectly tangential impact, as can be seen in Fig. \ref{fig:eccentricity}.}
        \label{fig:40AU_orbital_elements}
\end{figure*}

Taking a closer look at the trajectories shown in Fig. \ref{fig:40AUfidmodeltraj}, it is hard to visually determine whether there is circularisation or not. The pebbles with highly eccentric orbits in the system without an atmosphere, namely the red and blue ones, are slightly more contained around the planet and show fewer revolutions when the atmosphere is turned on, but an absolute reduction in eccentricity is hard to make out. To highlight the circularisation, Fig. \ref{fig:40AU_orbital_elements} shows the evolution of the instantaneous orbital elements during the final year before the impact of these orbits, represented in the same colours. The eccentricity of an orbit is calculated using
\begin{equation}
        e = \sqrt{1+\cfrac{2\epsilon(t) l_\mathrm{z}(t)^2}{(GM_\mathrm{p})^2}}    \label{eq:vis-vivaB},
\end{equation}
in which $\epsilon$ is the specific total energy of the orbit, given by
\begin{equation}
        \epsilon = \cfrac{v_\mathrm{pc}(t)^2}{2}-\frac{GM_\mathrm{p}}{R(t)} = -\frac{GM_\mathrm{p}}{2a_\mathrm{pc}(t)}    \label{eq:vis-vivaA}\\
,\end{equation}where $v_\mathrm{pc}$ is the pebble's velocity with respect to the planet, $a_\mathrm{pc}$ is the semi-major axis of the planetocentric orbit, and $R$ is the distance to the planet's centre.

The eccentricity decreases over time when the planet has an atmosphere, while the orbital elements remain more or less constant without an atmosphere. During the final phase of the orbits with an atmosphere, the eccentricity reaches a minimum of about zero after which it rapidly increases. This is paired with a sudden, rapid decrease in the semi-major axis, suggesting a phase of very strong orbital decay. It should be noted that the scaling of the x-axis changes from logarithmic to linear at $10^{-2}$ years, resulting in plotting artefacts at the interface. 

\begin{figure}[]
        \centering
        \includegraphics[width=0.925\linewidth]{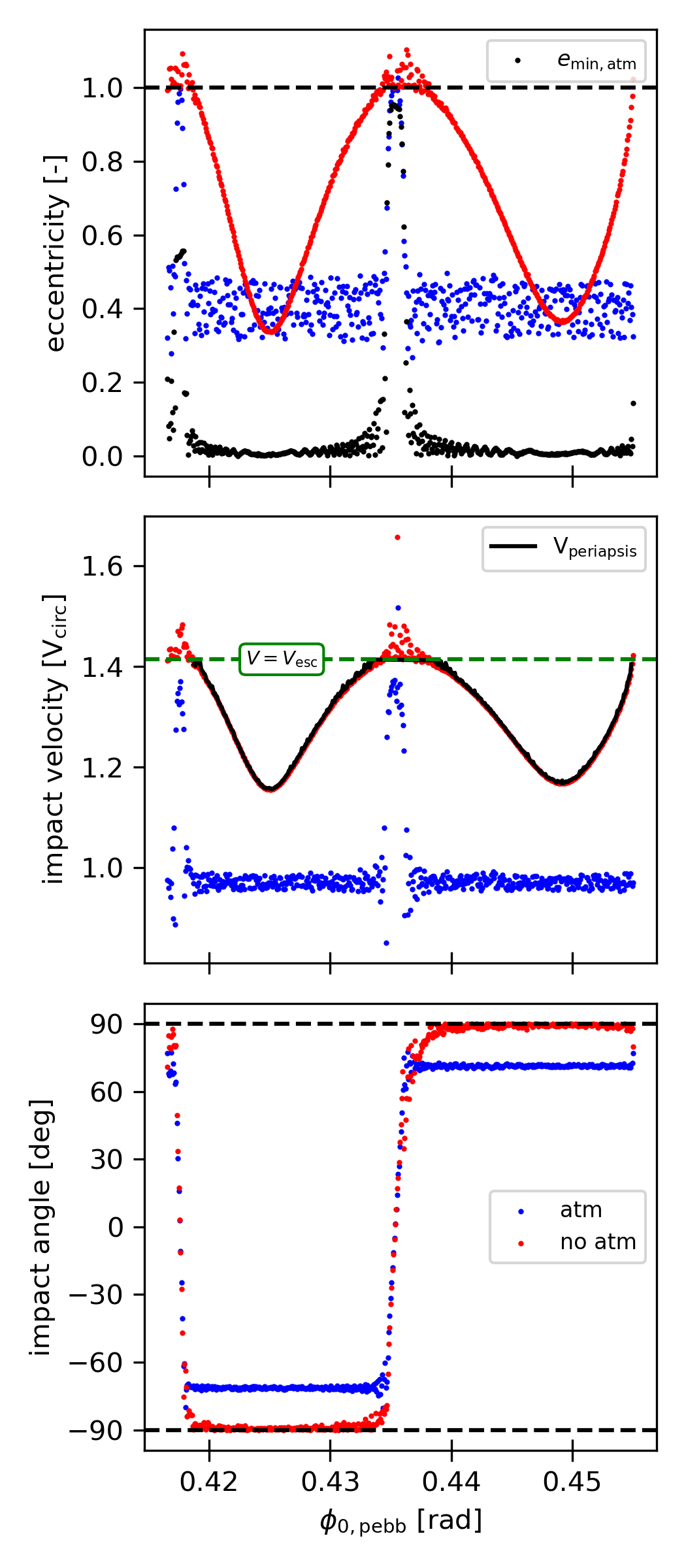} 
        \caption{Eccentricity, impact velocity, and impact angle of pebbles at the moment of collision with the planet as a function of starting position within the collision cross-section. The orbits in the presence of an atmosphere (blue) are significantly circularised compared to the orbits in the absence of an atmosphere (red), reducing the mean eccentricity from 0.708 to 0.432. However, the black dots in the top panel show the minimum eccentricity during the trajectories with an atmosphere, which is almost zero, indicating that at some point the eccentricity increases again. The impact velocity in the presence of an atmosphere is slightly below the circular velocity, while without an atmosphere it is equal to the velocity of the final orbits at periapsis. The latter, in combination with the almost perfectly tangential impacts of atmosphere-less pebbles, shows that the orbital decay without an atmosphere is minimal. With an atmosphere, the impact angle is at about $\pm 70^\mathrm{o}$. This, in combination with the slightly sub-Keplerian impact velocity, signifies a strong orbital decay in the final stage of the orbit right before impact. 
        }
        \label{fig:eccentricity}
\end{figure}

Further evidence of circularisation by the atmosphere in the fiducial model is shown in Fig. \ref{fig:eccentricity}. This figure shows the eccentricity of all pebbles at the moment of impact with the planet, along with the impact velocity and angle. The eccentricity of most pebbles that travelled through an atmosphere, henceforward the blue pebbles, is lower than that of the pebbles in the simulation without an atmosphere, the red pebbles. The atmosphere reduces the mean eccentricity at the moment of impact from 0.708 to 0.432. However, the minimum eccentricity of the pebbles during their full trajectory, represented by the black dots in the top panel of Fig. \ref{fig:eccentricity}, is almost zero. This shows that the eccentricity minima seen in Fig. \ref{fig:40AU_orbital_elements} are not specific to the selected pebbles, but that nearly all pebbles follow a perfectly circular orbit at some point in the atmosphere, after which the eccentricity starts to increase. The only exception to this behaviour is seen in the pebbles that, without an atmosphere, have an eccentricity above 1.0 and an impact velocity above the escape velocity. These pebbles are on hyperbolic trajectories with respect to the planet. The impact parameters of these trajectories are smaller than the planet's radius and therefore these pebbles directly impact the planet instead of orbiting it.

Another interesting point in Fig. \ref{fig:eccentricity} is that there is a strong variation and little correlation between the measured eccentricities of the blue pebbles. Meanwhile, the eccentricities of the red pebbles follow a very clear curve. The reason for this can be seen in the impact velocity and angle. For most of the domain, the impact velocity of the red pebbles is equal to the velocity at periapsis of the final orbit, given by
\begin{equation}
        V_\mathrm{p} = \sqrt{\frac{GM_\mathrm{p}}{a_\mathrm{pc}}}\sqrt{\frac{1+e}{1-e}},
\end{equation}
in which the $e$ and $a_\mathrm{pc}$ are calculated with Eqs. \ref{eq:vis-vivaB} and \ref{eq:vis-vivaA}. This, in combination with the almost perfectly tangential impact angle $\phi_\mathrm{impact}$ (50\% of pebbles: $|\phi_\mathrm{impact}|>89^\mathrm{o}$, 72\% of pebbles: $|\phi_\mathrm{impact}|>88^\mathrm{o}$) shows that the orbital decay without an atmosphere is minimal and that the impact occurs once the periapsis exactly coincides with the radius of the planet.

The blue pebbles, on the other hand, have an impact angle of approximately $\pm 70^\mathrm{o}$ and an impact velocity that is decreased slightly below the circular Kepler velocity. This is further evidence of the strong orbital decay in the final stages of the orbits in the presence of an atmosphere, which was observed in Fig. \ref{fig:40AU_orbital_elements}. During this decay, Eqs. \ref{eq:vis-vivaB} and \ref{eq:vis-vivaA}, which assume the orbital energy is constant throughout the orbit, are no longer valid, which is why the eccentricity at the moment of impact is not well defined either. The atmosphere in this fiducial model lies at the transition between tenuous atmospheres that slightly perturb the orbit of pebbles, and extremely dense atmospheres that absorb all angular momentum and orbital energy of the pebbles, after which the pebbles fall radially inwards. Therefore, this model shows signs of both circularisation due to orbital perturbations and of rapid orbital decay.\\

Finally, looking back at the atmosphere-less trajectories in Fig. \ref{fig:40AUfidmodeltraj} and comparing them with their deposited spin in Fig. \ref{fig:40AUfidmodelspin}, we see that the highest spin contributions, those close to $l_\mathrm{z,esc}$, come from highly eccentric pebbles, like the red and blue ones. These pebbles undergo the most significant change in deposited angular momentum when the atmosphere is turned on because of the circularisation. The cause of this circularisation is the fact that these orbits have their pericentre deeper within the planet's atmosphere than their apocentre. This means that the gas density is highest at the point of highest velocity and lowest at the point of lowest velocity. This drag difference causes the relative difference between the two velocity extremes to decrease, which can only be achieved through circularisation. On the other hand, a pebble that follows an orbit that spirals inwards more circularly, like the one in grey, is influenced by the atmosphere much less, since it experiences no density changes during a revolution around the planet. Their SAM is already close to Keplerian in the case without an atmosphere and is only slightly altered when the atmosphere is turned on.

\subsubsection{Atmospheric spin absorption profile}
\label{sec:fidmodel_atmospheric_absorption}
\begin{figure*}[t]
        \centering
        \includegraphics[width=0.8\linewidth]{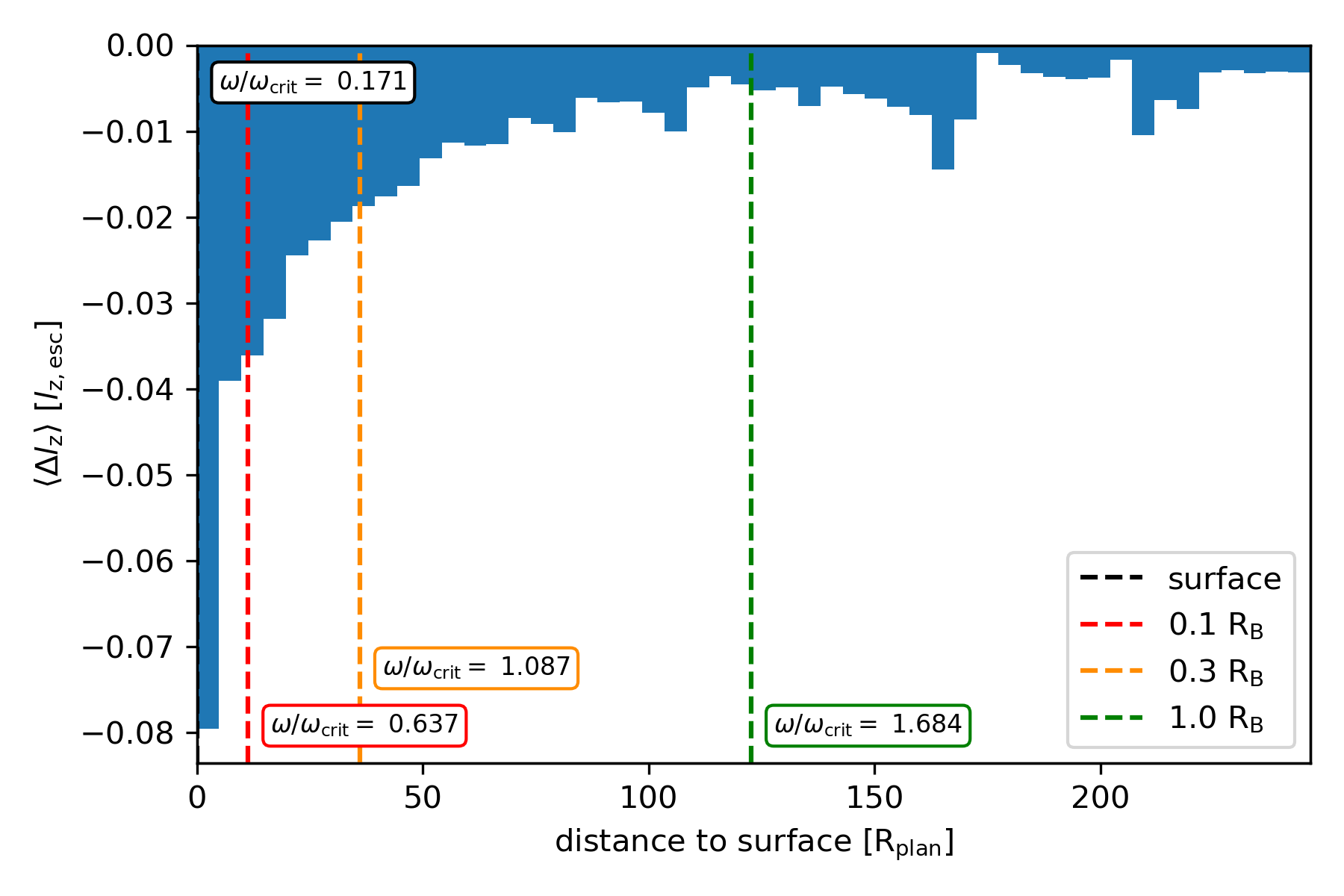}
        \caption{Mean change in angular momentum of the pebbles as a function of distance to the planet. The $\omega/\omega_\mathrm{crit}$ values show the total spin, expressed in the spin frequency normalised by the critical spin frequency of the planet, that the pebbles retain at distances smaller than the corresponding line. In total, the pebbles enter the atmosphere with approximately 1.684 $\omega_\mathrm{crit}$.  35.5\% of this spin is absorbed in the outer atmosphere at a distance larger than 0.3 Bondi radii, 26.7\% between 0.1 and 0.3 Bondi radii, 27.7\% in the inner atmosphere below 0.1 Bondi radii, and finally 10.2\% is deposited onto the planet.}
        \label{fig:dlz figure}
\end{figure*}

In conclusion, the atmosphere reduces the amount of spin that is transferred to the planet's surface by circularising the orbits of pebbles and reducing the impact angle. The rest of the spin is retained in the atmosphere itself. Figure \ref{fig:dlz figure} shows where in the atmosphere the angular momentum is lost. This profile of angular momentum loss as a function of distance to the planet's surface has been averaged over all 500 pebbles. The $\omega/\omega_\mathrm{crit}$ values show the averaged total spin, expressed in the spin frequency normalised by the critical spin frequency of the planet, that the pebbles retain at distances smaller than the corresponding line. In total, the pebbles enter the atmosphere with approximately 1.684 $\omega_\mathrm{crit}$. 35.5\% of this spin is absorbed in the outer atmosphere at a distance larger than 0.3 Bondi radii, 26.7\% between 0.1 and 0.3 Bondi radii, 27.7\% in the inner atmosphere below 0.1 Bondi radii and 10.2\% is deposited onto the planet.\\

There are three possible scenarios that describe what happens with the angular momentum after it has been deposited in the atmosphere. One possibility is that viscous forces transport all angular momentum to the surface of the planet. This would lead to a significant spin increase of planets with atmospheres compared to those without atmospheres. Another possibility is that all angular momentum is transported to the disc due to atmospheric recycling \citep{Ormel_2015_atm_rec_II}. In this case, the comparison between the two mean spin values in Fig. \ref{fig:40AUfidmodelspin} is valid and the atmosphere causes a reduction of the planetary spin of 26.9\%. Finally, a combination of both scenarios is possible, in which the innermost layers of the atmosphere are bound and have their spin transported to the planet's surface, while the outer layers are recycled and have their spin transported to the disc \citep{Cimerman_2017,Kurokawa_2018}.

These scenarios and their applicability to the results in this study are further discussed in Sect. \ref{sec:discussion-recycling}. The rest of the results in Sect. \ref{sec:results} are discussed under the assumption of full atmospheric recycling unless stated otherwise, leaving only the spin that is directly transferred to the planet's surface.\\

\subsubsection{Atmospheric influence on accretion efficiency}
\begin{figure}[t]
        \centering
        \includegraphics[width=0.85\linewidth]{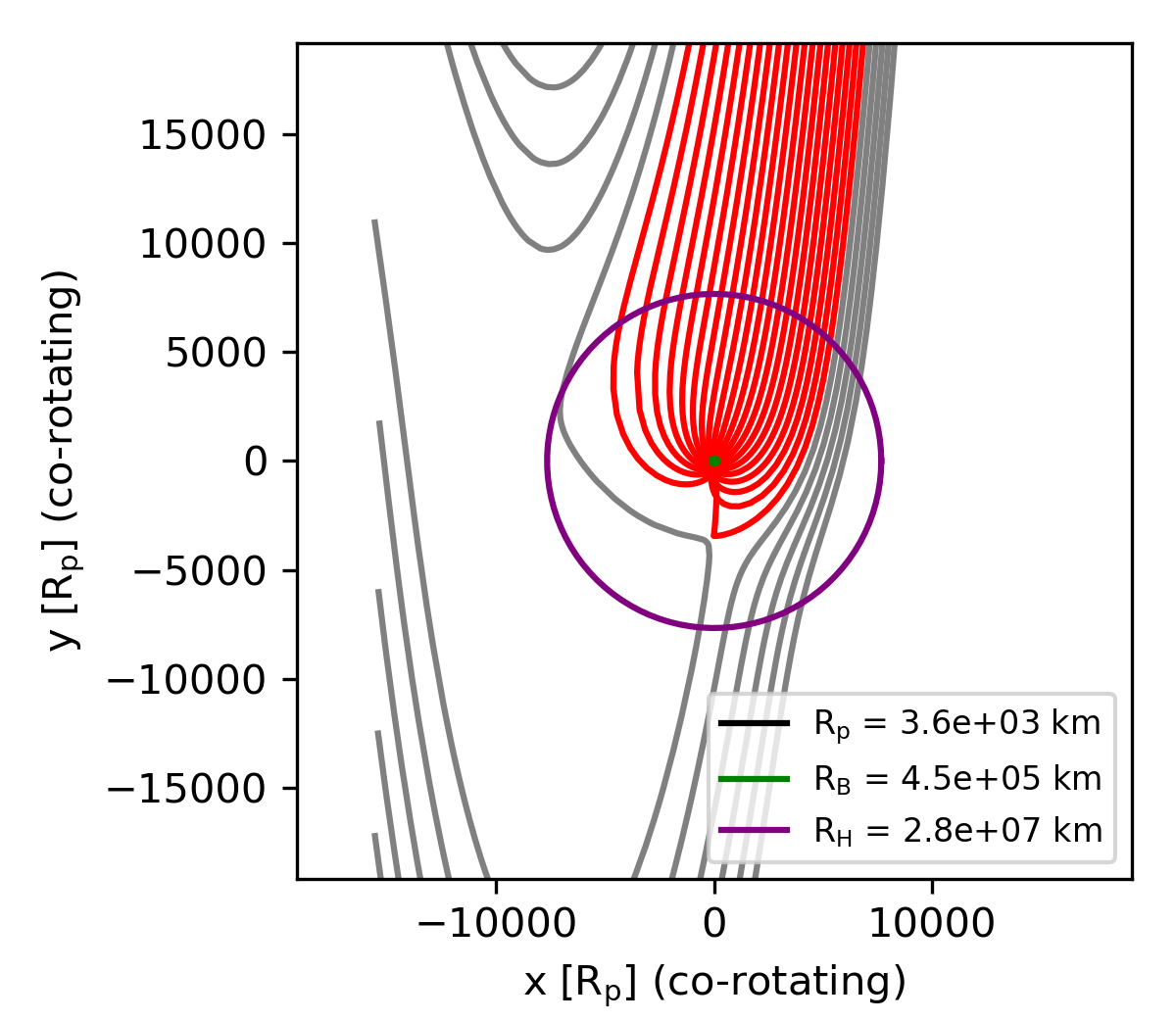}
        \caption{Zoomed-out image of the trajectories of the pebbles in the fiducial model. The pebbles fall in the settling regime, meaning that the encounter time is longer than the settling time. The pebbles `settle' in the gravitational field of the planet and spiral inwards. The atmosphere does not influence the accretion rate, since the collision cross-section is much larger than the atmosphere itself.}
        \label{fig:hillscale_traj}
\end{figure}

The atmosphere did not influence the accretion efficiency. The collision cross-sections for the situation with and without an atmosphere are identical. This is because the pebbles used in this study fall in the settling regime, in which pebbles settle in the gravitational field of the planet and slowly fall inwards \citep{Visser_2019}. The collision cross-section is determined by the planet's mass, not by any physical radius \citep{Ormel_2017_PA}. As can be seen in Fig. \ref{fig:hillscale_traj}, nearly all pebbles that enter the Hill sphere eventually accrete onto the planet, meaning that the cross-section is much larger than the atmosphere itself. Any pebble that enters the atmosphere was already on a collision course with the planet, and therefore the atmosphere cannot increase the number of collisions. This is true for all simulations in this study. The atmosphere might in fact reduce the accretion efficiency, especially for very small particles ($\tau_\mathrm{s}\ll 1$, through enhanced aerodynamic deflection \citep{Visser_2016, Ormel_2017_PA}, though this process is ignored in this study.   

\subsection{Spin transfer as a function of disc orbital radius and planet mass}
\label{sec:spin_afo_a}
\begin{figure*}
        \centering
        \includegraphics[width=\linewidth]{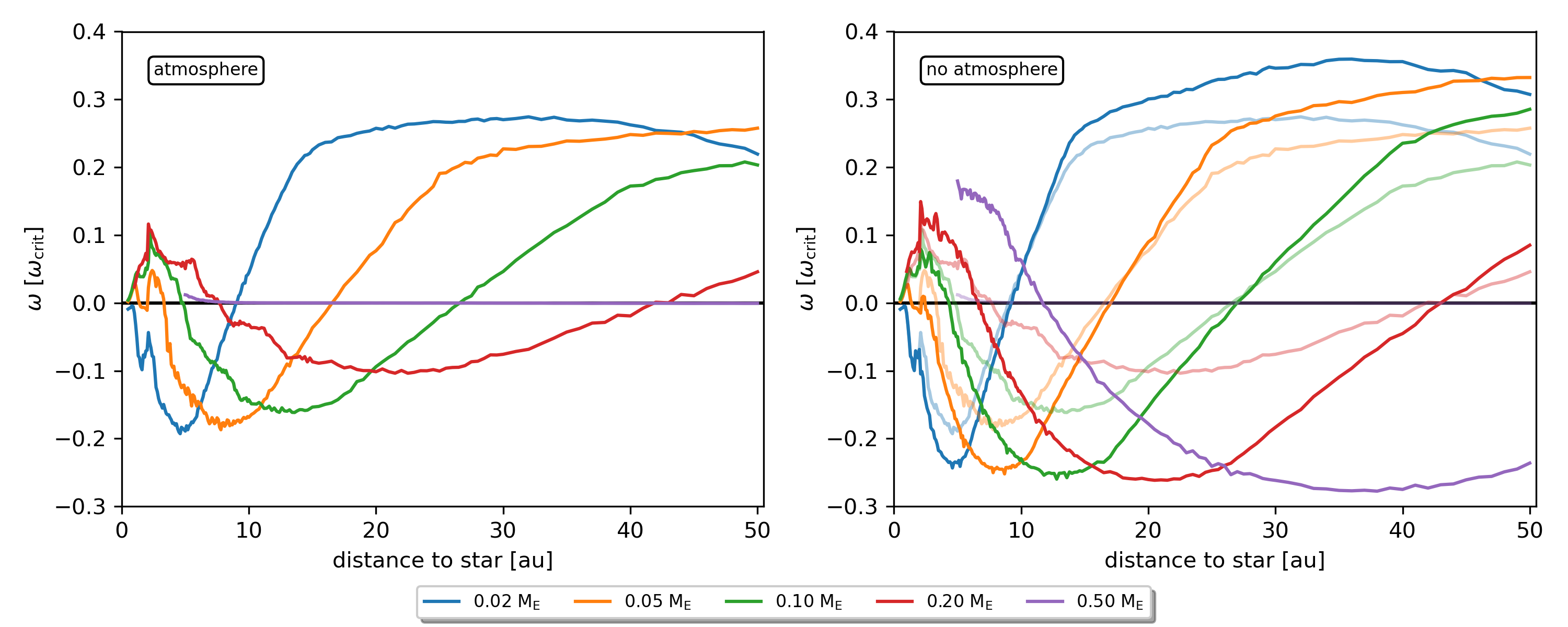}
        \caption{Mean spin angular frequency of the planet's surface as a function of disc orbital radius for different planet masses. The Stokes number of the pebbles in these simulations was 0.1. The left panel shows the spin of the planet's surface due to PA in the presence of an atmosphere. The right panel shows the spin in the absence of an atmosphere, with the spin with an atmosphere represented by transparent lines for direct comparison. In general, the atmosphere reduces the amount of spin that reaches the surface. The larger the planet, the stronger the spin reduction. For planets as small as 0.5 Earth masses, the atmosphere will absorb all spin. Additionally, the larger the planet, the larger the retrograde spin domain is, both with and without an atmosphere. Finally, as the spin crosses from retrograde to prograde, there is a small domain in which the atmosphere increases the amount of spin. This suggests a slight asymmetry between the damping of prograde and retrograde spin contributions.}
        \label{fig:spin_afo_a}%
\end{figure*}
Figure \ref{fig:spin_afo_a} shows the relationship between the spin a planet receives from PA and the distance from that planet to the star. It shows that in general, the atmosphere reduces the amount of spin that directly reaches the planet's surface, absorbing a significant portion of both the prograde and retrograde spin. The more massive the planet, the more spin is absorbed by the atmosphere. Since the Bondi radius is proportional to a planet's mass $M_\mathrm{p}$ (Eq. \ref{eq:Rbondi}), while the Hill radius is proportional to $M_\mathrm{p}^{1/3}$ (Eq. \ref{eq:Rhill}), the importance of the atmosphere during PA increases rapidly as the planets grow bigger. At about 0.5 Earth masses, the atmosphere becomes so dense, with the density at the planet's surface reaching 7700 times (10 au) to 36000 times (40 au) the disc density, that all angular momentum is absorbed in the atmosphere and no spin is transferred to the planet's surface.

\begin{figure*}
        \centering
        \includegraphics[width=\linewidth]{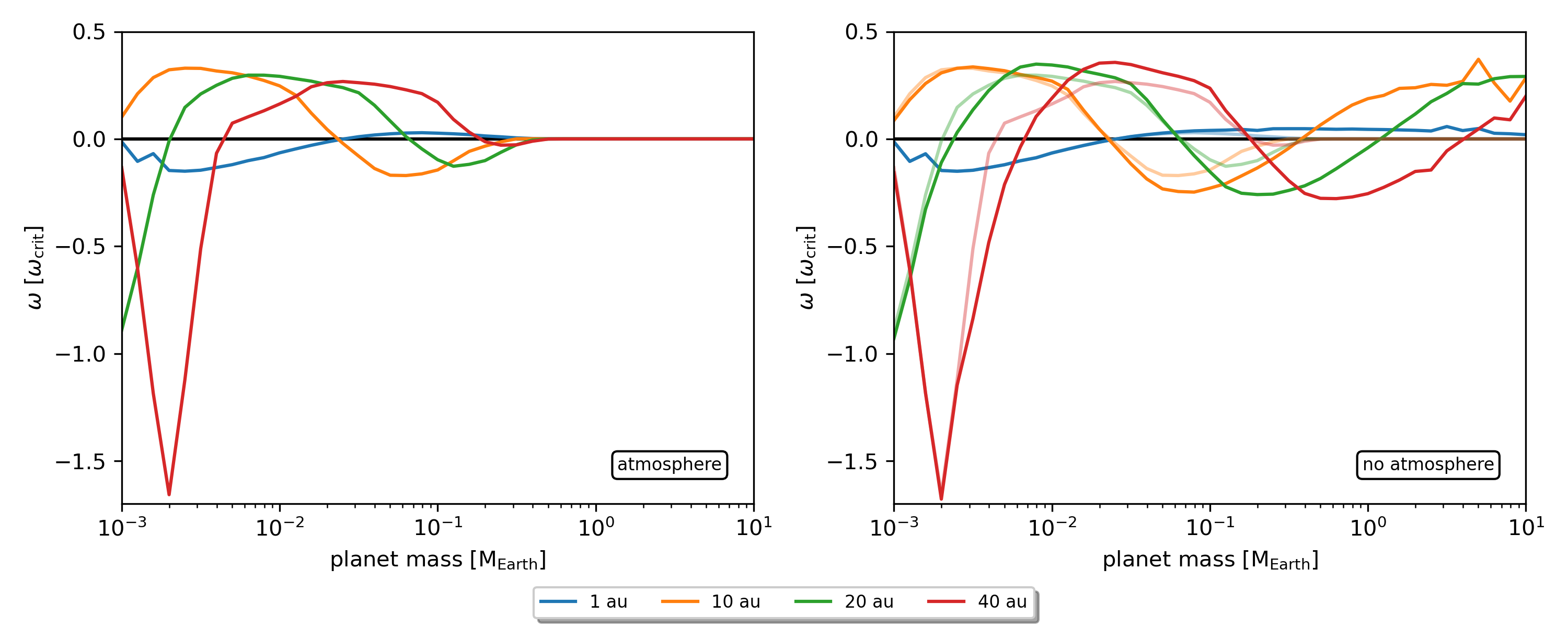}
        \caption{Spin of the planet due to PA as a function of planet mass. The more massive the planet, the less spin reaches the planet's surface. For planets of 0.5 Earth masses and above, the atmosphere absorbs all incoming spin. In the absence of an atmosphere, there is a large domain of retrograde spin for planets larger than ~0.1 - 0.2 Earth masses. An atmosphere could prevent this strong retrograde spin build-up and leave the planet with prograde spin at the end of its evolution.}
        \label{fig:spin_afo_M}
\end{figure*}

Another noteworthy point is that the region in disc-orbital-radius space in which retrograde spin dominates increases with increasing planet mass. Similar large retrograde regions can be seen in Fig. \ref{fig:spin_afo_M}, in which the spin on the planet's surface is studied as a function of planet mass for different distances to the star. Based on these figures, one might be tempted to argue that the ice giants would accumulate large amounts of negative spin, especially if the atmosphere would not play a part during PA. The fact that the atmosphere dampens the spin of massive planets almost completely could potentially reduce the influence of these retrograde contributions throughout the full evolution of the planet. However, it is important to note that the amount of spin a pebble transfers is dependent on three parameters, while these figures only show rough relations between two of these parameters. Pebbles with a different Stokes number will deposit a different, potentially prograde, spin \citep{Visser_2019}. In order to make predictions about the spin of the planets, the spin must be studied as a function of planet mass and Stokes number for a specific distance to the star.\\

Before we discuss those parameters, however, there is one final important feature visible in Fig. \ref{fig:spin_afo_a} worth mentioning. There seems to be an asymmetry in the damping mechanism of the atmosphere. As the spin crosses from retrograde to prograde, there is a small domain in which the atmosphere enhances the spin the surface receives. This is especially visible in the 0.05 and 0.1 Earth-mass simulations, at around 15 and 25 au, respectively. The bright and faint curves of these simulations in the right-hand panel of Fig. \ref{fig:spin_afo_a} do not intersect at exactly zero spin. Instead, the intersection seems to have a systematic offset of about +0.05 to +0.1 $\omega_\mathrm{crit}$, with a stronger offset for the 0.05 Earth-mass situation than the 0.1 Earth-mass situation. This shows that there is a domain in which the magnitude of the planetary spin with an atmosphere (faint curve) is larger than without an atmosphere (bright curve).

\begin{figure}
        \centering
        \includegraphics[width=\linewidth]{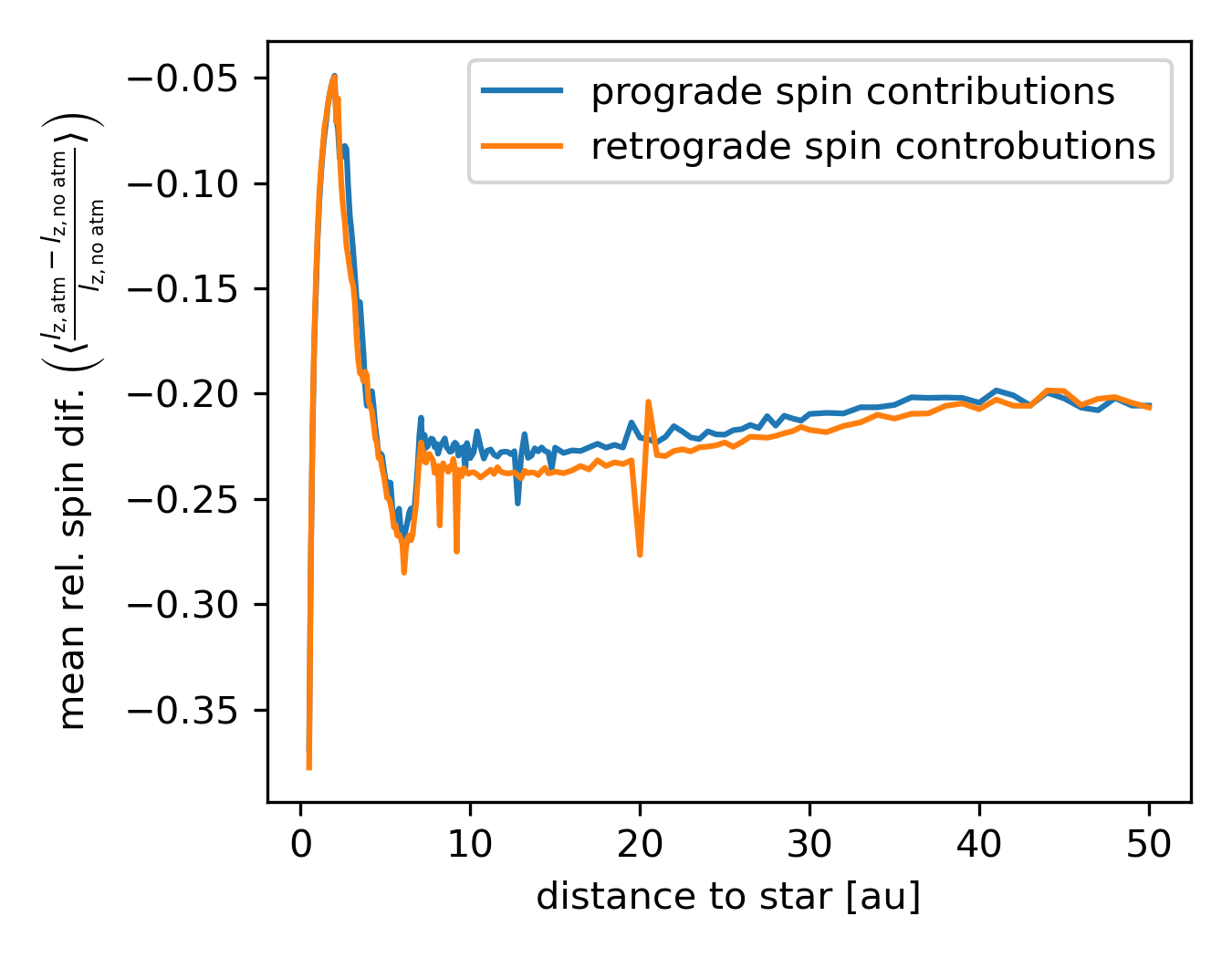}
        \caption{Comparison between the mean relative spin difference $\left\langle (l_\mathrm{z,atm}-l_\mathrm{z,no\ atm})/l_\mathrm{z,no\ atm} \right\rangle$ of prograde spin contributions and retrograde spin contributions for accretion on a 0.05 Earth-mass planet as a function of distance to the star. Retrograde spin contributions are more strongly dampened than positive spin contributions.}
        \label{fig:dampingasymmetry}
\end{figure}

Since the spin enhancement by the atmosphere only occurs when the prograde spin preference without an atmosphere is small, the atmosphere likely dampens retrograde spin slightly more strongly than prograde spin. The enhancement occurs in a regime in which the individual prograde and retrograde contributions are both strong ($>l_\mathrm{z,Kepler}$), but in balance, nearly perfectly cancelling each other out. A slight asymmetry in the damping mechanism could then result in a relative boost in prograde spin that is strong enough to result in a nett spin increase. If the asymmetry is small enough, the effect would be unnoticeable when there is a significant spin preference, since the amount of average damping is much stronger than the damping asymmetry. Figure \ref{fig:dampingasymmetry} confirms the asymmetry, showing that in the relevant domain (between 10 and 20 au), an individual retrograde spin contribution is on average dampened 1\% more than a prograde spin contribution. The source of such asymmetric damping could most likely be found in the shear and/or Coriolis force when evaluating the situation in a planetocentric frame. Both of these forces act in opposite directions for pebbles orbiting in a prograde fashion compared to those orbiting in a retrograde fashion. The spin enhancement due to the atmosphere is even stronger in Fig. \ref{fig:spin_afo_M} for small planets ($\sim 0.005\ {\rm M}_\Earth$) further out in the disc (20 and 40 au), suggesting that the mass of the planet and thereby the accretion regime plays an important part as well.

\subsection{Spin transfer as a function of planet mass and Stokes number}
\label{sec:spin_afo_M_and_Stk}
\begin{figure*}[t]
        \centering
        \includegraphics[width = 0.9\linewidth]{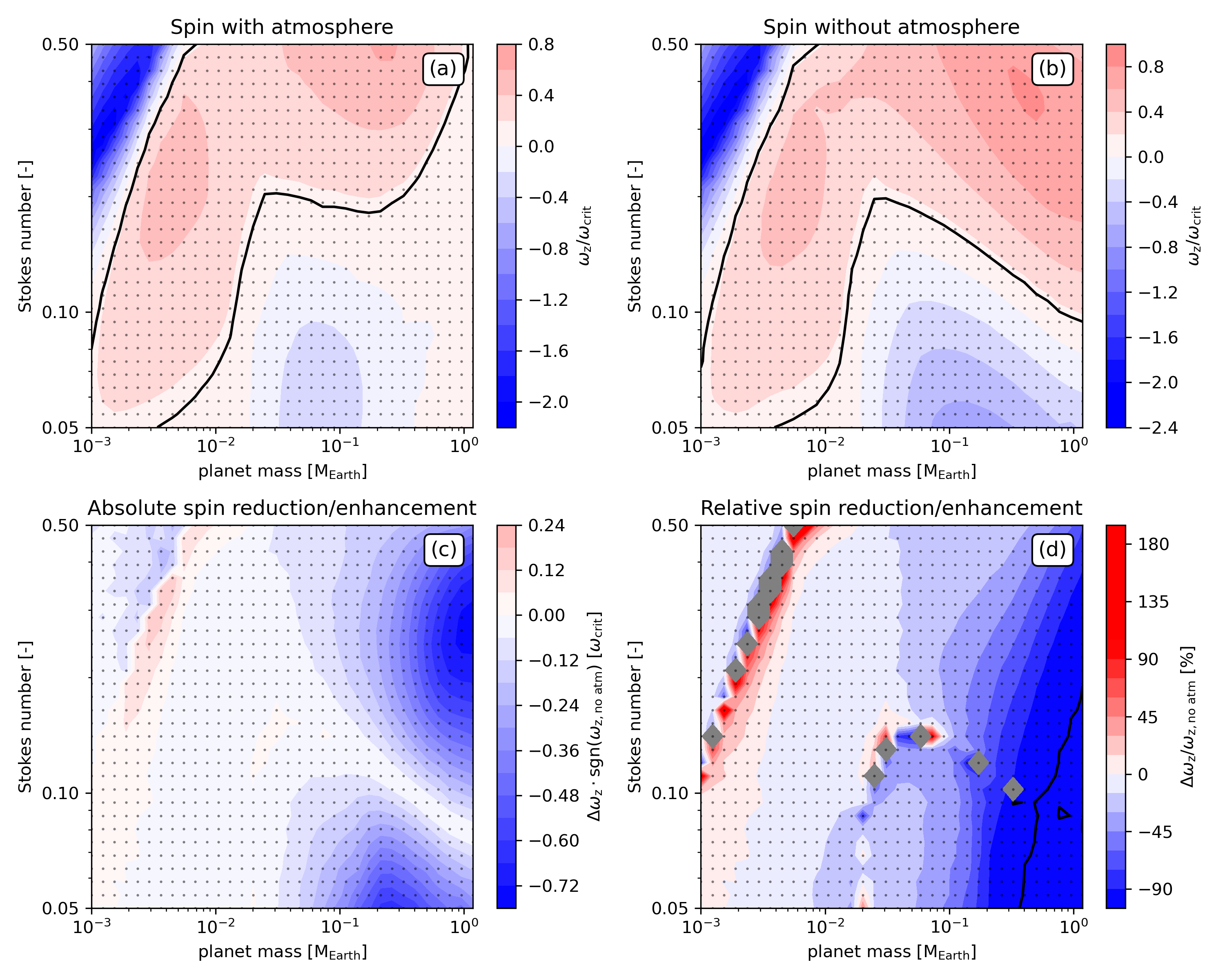}
        \caption{Spin deposited directly on the planet's surface as a function of Stokes number and planet mass for a planet at 10 au. Panel (a) shows the spin in the presence of an atmosphere and panel (b) when there is no atmosphere. To give an impression of how these spin values relate to those measured in the Solar System, the spin value of Neptune ($\omega_\mathrm{z} \approx 0.16\omega_\mathrm{crit}$) is indicated in panels (a) and (b) using a black contour. In panel (c) the absolute reduction (blue) or enhancement (red) of the spin due to the atmosphere is plotted. In this panel, the absolute difference has been multiplied by the sign of the spin without an atmosphere so that a reduction brings the spin closer to zero, while an enhancement moves the spin farther from zero, irrespective of whether the spin is prograde or retrograde. Finally, panel (d) shows the percentage difference between the two models. The black contour in this plot indicates the region in which 100\% of the spin is absorbed by the atmosphere. The colour scale in panel (d) has been cut off at -100\% and +500\%. Data points outside this range, which are artefacts due to the spin without an atmosphere becoming zero, have been masked (grey).}
        \label{fig:spin_Stk_vs_M}
\end{figure*}

Figure \ref{fig:spin_Stk_vs_M} shows the spin rate of a planet's surface as a function of the Stokes number of the pebbles and the planet's mass for a planet at 10 au, as well as the absolute (c) and relative (d) spin reduction and enhancement due to the atmosphere. This figure shows that even though the results in Figs.  \ref{fig:spin_afo_a} and \ref{fig:spin_afo_M} predict a strong retrograde spin for massive planets, these planets still receive prograde spin through pebbles with higher Stokes numbers. Figure \ref{fig:spin_Stk_vs_M}b is in strong agreement with the planar (2D) results in \citet{Visser_2019}. The large prograde spin region in this plot predicts values that can match and even exceed the current spin of the large bodies in the Solar System.\\

For comparison with the kind of angular momentum present in the planets in the Solar System, the spin of Neptune ($\omega_\mathrm{z} \approx 0.16\omega_\mathrm{crit}$) is shown in Figs. \ref{fig:spin_Stk_vs_M}a and b (solid black contour). Without an atmosphere, a planet can accumulate prograde spin equal to that of Neptune throughout the full evolution from a 0.001 Earth-mass planetesimal to a 1.0 Earth-mass planet, initially by accreting mainly smaller pebbles ($\tau_\mathrm{s}< 0.1$) and during later stages by accreting larger pebbles ($\tau_\mathrm{s} \gtrsim 0.1$). This regime of strong prograde spin build-up extends to at least planet masses of 6 M$_\Earth$, as was demonstrated by \citet{Visser_2019} in Fig. 3 of their paper. With an atmosphere, however, the planet will no longer receive any spin directly from PA once it reaches a mass of 0.5 to 1.0 Earth mass, with 100\% of the spin being absorbed, as indicated by the black contour in Fig.\ref{fig:spin_Stk_vs_M}d. In order for the planet to maintain a (partially) pebble-induced spin, significant prograde spin from PA must be stored in the primordial atmosphere and maintained there for a longer time, spinning up the atmosphere and transforming the system from the static extreme in our study to the rapidly rotating regime from \citet{Takaoka_2023}.

Naturally, the comparison between the planets in this study and Neptune is not completely fair, since these planets would still have a long evolution ahead of them, growing by a factor of 20 in mass, before they reach the mass of Neptune. The final spin state is determined in the final stage of this growth. Moreover, an ice giant might not have as clear an interface between the solid core and the atmosphere as the terrestrial-like planets in these simulations. Nevertheless, the results do give a first-order idea of the expected spin rates and the influence the atmosphere has on acquiring these spin rates, possibly proving important in predicting spin values for Earth-like and super-Earth-like exoplanets.\\

On a separate note, looking at the absolute and relative spin reduction and enhancement in Figs. \ref{fig:spin_Stk_vs_M}c and \ref{fig:spin_Stk_vs_M}d, there are systematic regions in the parameter space in which the atmosphere increases the spin on the planet's surface. These regions do not all fall in low-spin domains, showing that the asymmetry in the damping mechanism discussed in the previous section is significantly stronger for certain combinations of Stokes numbers and planet masses.\\

\subsection{Spin stored in the atmosphere}
\begin{figure*}
        \centering
        \includegraphics[width=\linewidth]{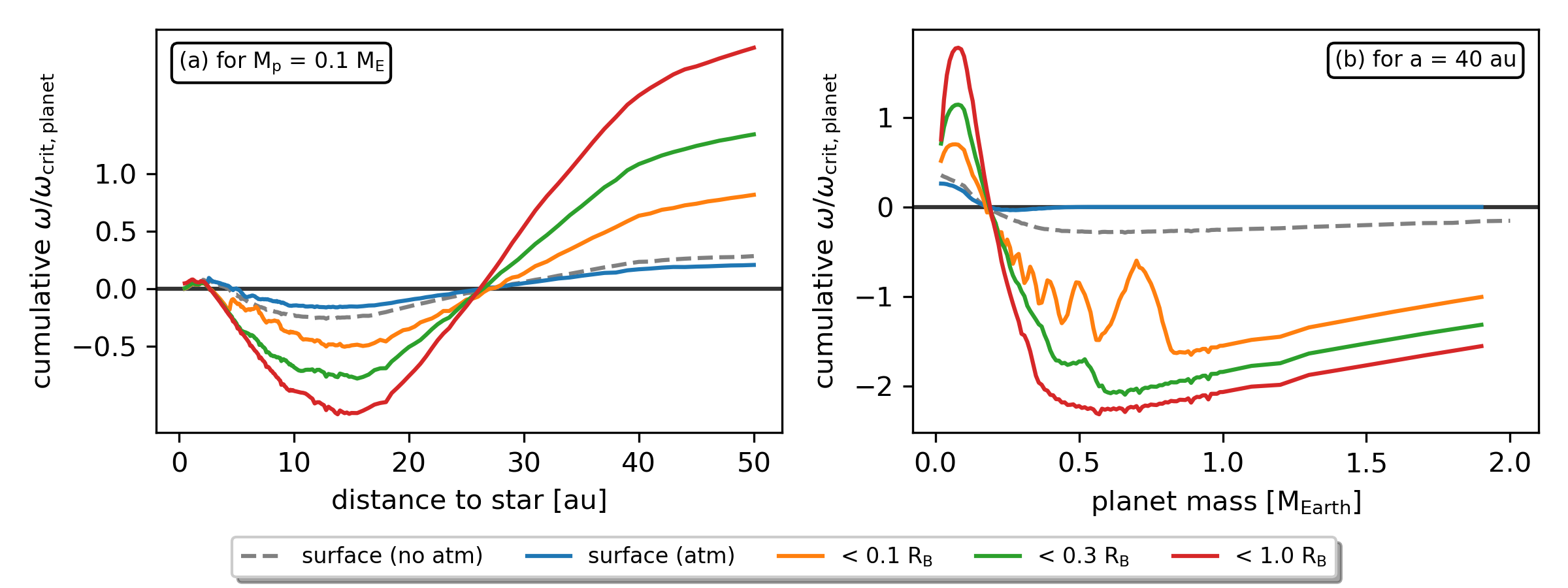}
        \caption{Cumulative spin stored within different boundaries in the atmosphere as a function of disc orbital radius (left) and planet mass (right). The spin values represent the spin rate of the planet if all angular momentum stored at altitudes below the boundaries, including the angular momentum directly deposited on the planet, was transported to the planet's surface. The spin stored within 0.1 Bondi radii already exceeds the accumulated spin in the atmosphere-less simulations.}
        \label{fig:stored_spin}
\end{figure*}

As was shown in Fig.\ref{fig:spin_Stk_vs_M}, for the planet to be able to accumulate Neptune-like spin from PA throughout its full evolution, significant quantities of prograde spin must be stored in the innermost regions of the atmospheres of massive planets and transferred to the surface. If most of the angular momentum is stored in the outer regions of the atmosphere, atmospheric recycling might prevent the spin from ever reaching the planet. To get an idea of the amount of spin that is stored in the atmosphere, the analysis described in Sect. \ref{sec:fidmodel_atmospheric_absorption} was performed for a planet of 0.1 Earth masses as a function of distance to the star, as well as for a planet at 40 AU as a function of mass. The results can be seen in Fig.\ref{fig:stored_spin}.

This figure shows that though the surface of a planet with an atmosphere receives less spin than one without, the spin contained in the inner 10\% of the atmosphere is equal to or exceeds the spin that reaches the planet without an atmosphere. The spin stored in the atmosphere is always in the same direction as the spin on the surface and all layers of the atmosphere cross from retrograde to prograde spin at the same time, showing that the different layers in this model atmosphere do not significantly influence the balance between prograde and retrograde spin. This might not be true for a (differentially) rotating atmosphere.

\section{Discussion}
\label{sec:discussion}
In this section the limitations of the results are discussed. We start with the main limitation to the direct interpretation of the results, namely that the transport of angular momentum in the atmosphere after it has been extracted from the pebbles, has been ignored. The spin that is directly deposited onto the surface has been measured, as well as the average torque on the atmosphere as a function of altitude, but that is not enough to determine how much angular momentum the planet will have once PA ceases. This section is followed by a short discussion about the relevance of the static atmosphere used in this study. Afterwards, we discuss other caveats, as well as make recommendations for further research.\\

\subsection{Atmospheric spin transfer}
\label{sec:discussion-recycling}
The results from Figs. \ref{fig:spin_afo_a}, \ref{fig:spin_afo_M}, and \ref{fig:spin_Stk_vs_M} show the spin frequency of the planet's surface ignoring all angular momentum that did not directly reach the planet. To truly predict the spin of the planet, even in the extreme case that the atmosphere starts without rotation as is assumed in this study, one would have to include angular momentum exchange between the planet and the atmosphere. This analysis was beyond the scope of this paper, requiring hydrodynamical simulations rather than the N-body simulations used here. However, based on hydrodynamical simulations of other authors we can analyse a few different scenarios. These scenarios were already briefly mentioned in Sect. \ref{sec:fidmodel_atmospheric_absorption}.\\

In the first scenario, all spin that enters the atmosphere is eventually transferred to the planet. In this case, planets with an atmosphere acquire far more spin than atmosphere-less planets, since angular momentum that is normally lost to the disc is now stored in the atmosphere and contained around the planet. However, this scenario is extreme and is not realistic. As can be seen in Fig.\ref{fig:dlz figure}, when dividing the atmosphere into concentric shells of equal thickness, the inner shells receive more angular momentum than the outer shells. Combined with the fact that outer shells have a greater moment of inertia, the inner layers of the atmosphere will be spun up more rapidly than the outer layers, leading to a negative velocity gradient in the radial direction. Ignoring the feedback from this induced rotational velocity of the atmosphere on the pebbles' spin deposit, viscous forcing between layers, therefore, dictates that most angular momentum is transported outwards, back to the disc, rather than inwards to the planet. There will therefore never be a situation in which all angular momentum in the planet-atmosphere system ends up on the surface.\\

On the contrary, it is more likely that all atmospheric angular momentum is transported to the disc, especially for smaller planets. Not only is the angular momentum transported outwards by friction between the atmospheric layers as discussed before, but the atmosphere as a whole is recycled on a relatively short timescale as well. Using hydrodynamical simulations with inviscid, isothermal gas forming a primordial atmosphere around embedded planets, \citet{Ormel_2015_atm_rec_II} find that the atmospheres are open systems. Gas enters the Bondi sphere at high latitudes and flows out through the midplane or vice versa. Replenishment is primarily a surface effect, with decreasing efficiency when more mass is located near the interior, as is true for isothermal atmospheres. The gas at 0.2 $R_\mathrm{B}$ is recycled on a timescale of $\sim \Omega_\mathrm{k}^{-1}$, with $\Omega_\mathrm{k}$ the orbital frequency of the planet. Gas at higher altitudes is replenished faster and at lower altitudes recycling is slower, though not significantly. \citet{Ormel_2015_atm_rec_II} conclude that the full envelope is only temporarily bound. The full replenishing timescale assuming a shear-dominated velocity profile is given by \citep{Ormel_2015_atm_rec_II}
\begin{equation}
        t_\mathrm{replenish} \sim 10^4\ \mathrm{yr}\ \frac{\chi_\mathrm{atm}}{f^*_\mathrm{cover}}
        \left(\frac{M_\mathrm{p}}{M_\Earth}\right)^{-2} 
        \left( \frac{a}{1\ \mathrm{au}}\right)^{2.75},
\end{equation}
in which $\chi_\mathrm{atm}$ is the mass fraction of the atmosphere $M_\mathrm{atm}/M_\mathrm{p}$ and $f^*_\mathrm{cover}$ is the fraction of gas streams that reach a minimum depth of $r = r^*$. The $r^*$ corresponds to the shell where most atmospheric mass resides, meaning where $\rho(r)r^3$ peaks. For fully convective atmospheres as used in this study $r^*\sim R_\mathrm{B}$ and $f^*_\mathrm{cover}\sim 1$. Using these values and the parameters in our study, we find that the replenishing timescale is approximately 6.5 years for a planet at 1 au and 1700 years for a planet at 40 au. The mass of the planet has no significant influence on this timescale. For comparison, the time it takes for a pebble to settle is given by
\begin{equation}
        t_\mathrm{sett}=\frac{b^3}{GM_\mathrm{p}t_\mathrm{s}},
\end{equation}
with $t_\mathrm{s}$ the stopping time of the pebble and $b$ the impact parameter, given by the quadratic equation \citep{Ormel_2010}
\begin{equation}
        b^2\left(v_\mathrm{hw}+\frac{3}{2}\Omega_\mathrm{k}b \right)=4GM_\mathrm{p}t_\mathrm{s}.
\end{equation}
Here $t_\mathrm{sett}$ is 20 to 40 times smaller than the replenishment timescale for planets at 1 and 40 au, respectively, and pebbles with a Stokes number of 0.1. However, the recycling timescale should be compared to the time it takes to significantly spin up the atmosphere and for the induced friction between the planet and the atmosphere to transport the spin to the planet. This will depend on the accretion rate, as well as on the type of surface of the planet. On the one hand, for low accretion rates and a sharp interface between the surface and the atmosphere, nearly all angular momentum will likely be transported to the outer layers of the atmosphere since the angular momentum transport within the atmosphere along the negative rotational velocity gradient is much more efficient than the angular momentum exchange between the gas and the solid core. The atmospheric gas could then be recycled by gas from the disc before the atmosphere's velocity profile has had time to significantly evolve, thereby maintaining the gradient.

On the other hand, for high accretion rates, the atmosphere could be spun up much faster. If the interface between the core and the metal-free atmosphere is not sharp, but instead there is a layer of metal-rich vapour in between \citep{Brouwers_2018}, angular momentum exchange between the atmosphere and surface is much more efficient due to metals raining out of the atmosphere. This could leave the core with higher spin values than in the atmosphere-less situation. Further investigation is required to determine where the transition between these two regimes lies.\\

As for the applicability of the atmospheric recycling simulations of \citet{Ormel_2015_atm_rec_II} to our planets, the planet must be embedded, meaning that the planet radius is smaller than the Bondi radius, which in turn is smaller than the scale height of the disc $H$. This condition is valid for every planet in this study. As for the planet mass, \citet{Ormel_2014_atm_rec_I} performed their analysis on planets with a maximum $R_\mathrm{B}/H$ scale of $10^{-2}$. In our simulations, this corresponds to planets of 0.1 $M_\Earth$ at $a_\mathrm{p} \ge 1$ au, or 0.5 $M_\Earth$ at $a_\mathrm{p} \ge 5$ au, which is valid for the results in Fig.\ref{fig:spin_afo_a}. It is therefore likely that the results with an atmosphere in Figs. \ref{fig:spin_afo_a} and \ref{fig:spin_Stk_vs_M} show the true spin frequency a planet with a static atmosphere could accumulate from PA, at least for low accretion rates. This also means that planets with a mass $>0.5\ M_\Earth$ acquire no spin from PA.\\

Nevertheless, this conclusion still relies on the atmospheric model. \citet{Ormel_2015_atm_rec_II} used an inviscid, isothermal atmosphere. \citet{Popovas_2018} performed hydrodynamical simulations of adiabatic atmospheres, which are more applicable to the results of our study. The authors find similar non-circular streamlines deep within the atmosphere that are connected to the disc. There is no dynamic boundary between the bound atmosphere and the gas of the disc. This further supports the idea of full atmospheric recycling, though \citet{Popovas_2018} only studied Mars- and Earth-sized embryos at $1-1.5$ au. On the other hand, \citet{Kurokawa_2018} find that the inflow of the high-entropy disc gas into the cooled deep atmosphere is prevented by a buoyancy barrier that limits recycling in non-isothermal simulations. This buoyancy barrier is induced by a positive entropy gradient in the atmosphere. In non-isothermal radiation-hydrodynamics simulations, an isolated inner envelope appears as well, especially for high-mass planets. \citet{D'Angelo_2013} find an interface between bound and non-bound gas at $\sim 0.4\ R_\mathrm{B}$ for planets of $5-15\ M_\Earth$ at $5-10$ au. \citet{Cimerman_2017} find an inner core at $\sim 0.25\ R_\mathrm{B}$ within which the streamlines were more circular. They analysed planets with $R_\mathrm{B}/H$ parameters of 0.04, 0.38, 0.75 and 1.9, which corresponds to planet masses of $0.25-10.0\ M_\Earth$ at 1 au or $3.5-160\ M_\Earth$ at 40 au.\\

These studies show that there is a third possible scenario describing what happens with the spin in the atmosphere. In this final scenario, the angular momentum stored in the inner atmosphere can be transferred to the planet, while the angular momentum in the outer atmosphere is transferred back to the disc. This is especially likely for massive planets that in the current model accumulate no spin. For these planets, the pebbles lose most of their angular momentum at intermediate altitudes ($0.1\ R_\mathrm{B}<0.3\ R_\mathrm{B}$), within the isolated core. Part of this angular momentum is transported inwards, while the rest is transported outwards. The exact spin of the planet after PA thus depends on the balance between inward and outward transported angular momentum and on the size of the isolated core, as well as the assumed conditions of the atmosphere and surface. In the current, adiabatic atmosphere this scenario is not applicable since the isolated core is not formed. Further research is required to point out whether this transport could be a viable source of planetary spin when the atmosphere is neither isothermal nor adiabatic.

\subsection{The relevance of the static atmosphere}
\label{sec:discussion-atmrot}
In this study, we analysed the extreme case of PA in the presence of a static atmosphere. This scenario is in contrast to a recent study by \citet{Takaoka_2023}, who used hydrodynamic simulations to calculate a gas flow profile around a planet, which they then used in the numerical integration of 3D pebble orbits. A key difference between their model and ours is that the envelopes in their study are commonly prograde rotating due to the Coriolis force. Because of this rotation, originating from the connection between the planets and the dynamics in the disc, the atmospheres form an infinite source of positive angular momentum for the pebbles. Meanwhile, our planet-atmosphere systems are local, isolated systems, only connected to the disc through the boundary conditions of the temperature, pressure and density profiles. Instead of a source, the atmospheres are an angular momentum sink. 

The result is that, whereas we report significant orbit circularisation and a reduction in spin build-up, \citet{Takaoka_2023} find that planets receive prograde spin irrespective of the planet's mass, orbital radius or the pebbles' Stokes number. 
Since in their simulations, the envelopes become thicker and their prograde rotation stronger with increasing planet mass or orbital radius, massive planets further out in the disc can spin up to their break-up spin, leading to a rotationally induced isolation mass beyond which the planet cannot grow through PA. This is the exact opposite of our situation, in which those same massive planets acquire no spin from PA. 

However, similarly to us, \citet{Takaoka_2023} have ignored the back-reaction from the pebbles on the gas, meaning that the motion of the gas is unaltered by the interaction with the pebbles. Even though \citet{Takaoka_2023} assume different initial conditions for the planet-atmosphere systems, the dynamics included in the numerical simulations are the same as in our models. In a sense, they studied the other extreme to our static atmosphere: an atmosphere whose rotation is fully coupled to the dynamics in the disc, specifically to the Coriolis force. Both of the atmospheres, however, are fully decoupled from the incoming pebbles. \\

A specific example case in which a static atmosphere might be applicable is a system with a massive planet and a high incoming flux of small pebbles. As can be seen in Fig.\ref{fig:spin_Stk_vs_M}b, massive planets without atmospheres acquire strong retrograde spin states from pebbles with small Stokes numbers. Figure \ref{fig:stored_spin} shows that under those conditions, significant quantities of retrograde spin are absorbed by the atmosphere, even in the static case in which prograde and retrograde angular momentum are extracted with similar efficiencies. Given this large negative spin build-up in the atmosphere, one can imagine there is a situation in which the back-reaction from the pebbles on the gas cancels out the rotation from the Coriolis force and the atmospheric recycling, especially since the pebble mass accretion rate per year can exceed the total atmospheric mass, at least for the adiabatic atmospheres used in this study. A complete study of the motion of the gas, including the back-reaction from the pebbles, is required to determine if and when the static extreme is more appropriate than the rotating extreme. Either way, through the difference between our results and those of \citet{Takaoka_2023}, we show that the assumed initial atmospheric spin state has a very large effect on the calculated planetary spin states and that further investigation into the initial conditions is required.   

\subsection{Other caveats}
The simulations in this study aimed to determine the order of magnitude of the effect a static primordial atmosphere can have on spin transfer during PA and the shape of the pebble orbits. The relevance of this static model was briefly discussed in the previous section. However, in order to efficiently study a large parameter space with limited computing resources, a few important effects of PA and fluid dynamics were ignored during the simulations, aside from the angular momentum transfer within the atmosphere as discussed in Sect. \ref{sec:discussion-recycling} and the initial rotation state of the atmosphere and the back-reaction of the pebbles on the gas as discussed in Sect. \ref{sec:discussion-atmrot}.\\

The most important effects that were ignored relate to the motion of the gas. The planet was placed over a background of unperturbed gas. The velocity of the gas around the planet is purely governed by the headwind and shear velocity. The shear is an effect in a local frame co-rotating with the planet resulting from the decrease in circular orbital velocity as the orbital radius increases. Because of the shear, gas streamlines exterior to the planet's orbit point downwards, while gas interior to the planet flows upwards when looking at the planar model from above. The co-rotation line, in which the gas has the same velocity as the planet, is shifted slightly interior to the planet's orbit due to the constant headwind. However, in the simulations, the gas only exists for the pebbles and it flows right through the planet as though it does not exist.

In reality, the gas flows around the planet, changing the direction of the gas velocity from purely azimuthal around the star to a more complicated, potentially turbulent state around the planet. This would alter the amount and direction of the drag the pebbles experience, which in turn alters the trajectory of the pebbles. To incorporate this flow around the planet, one could locally modify the gas velocity using a prescription of either an inviscid potential flow or a viscous Stokes flow, both of which are described by \citet{Batchelor_1967}. The differences between these two models are small. The Stokes flow has an additional $\propto 1/r$ term in its description that makes its influence reach farther into the disc. Another difference is that the Stokes flow at the planet's surface is zero, while the potential flow at the surface is non-zero and in the azimuthal direction. This mainly influences low Stokes number particles (Re $\lll$ 1) that are completely coupled to the gas. These pebbles experience contrasting effects depending on which flow is assumed. They either have their interaction time increased due to the low-velocity Stokes flow close to the planet, increasing their accretion rate, or they experience a strong aerodynamical boundary due to the potential flow around the planet, preventing them from accreting. For larger pebbles and larger planets (>500 km), however, the difference between the two models is negligible \citep{Visser_2016}.\\

The main problem with these flow models is that they assume a fixed boundary that the gas cannot pass, namely the planet's surface. This description is therefore not applicable to the flow around a planet with an atmosphere. The streamlines will partially penetrate the atmosphere and be more and more deflected the deeper they reach. To get a sense of these flow patterns, hydrodynamical simulations are required.\\

The atmosphere used in the simulations was mainly a density and temperature model. By reducing the headwind for pebbles deep in the atmosphere, the gas was more `bound' to the planet, though shear still played a part. Further investigation is required to determine the influence of the shear on the angular momentum of the pebble and to determine whether a reduction in shear is appropriate. Nevertheless, the influence is likely limited. Since most of the analysis is performed at large distances from the star, the shear within the atmosphere is very small. Even for the large 0.5 Earth-mass planets, the gas velocity difference due to shear between the points in the atmosphere closest to and furthest from the star is only 0.55 to 1.09 m/s at 40 au and 20 au, respectively. For a 0.1 Earth-mass planet, this is 0.11 to 0.22 m/s. In the inner atmosphere, the difference is even less.\\

As mentioned in Sect. \ref{sec:atmosphericmodel}, we have assumed the atmosphere consisted of a single adiabatic layer. In doing so, we have ignored the two other atmospheric regions as proposed by \citet{Brouwers_2020}. In the innermost regions of the atmosphere, a convective heavy vapour layer could form. In this layer, the temperature is so high that the pebbles thermally ablate. If for simplicity's sake we assume that a pebble fully evaporates the moment its surrounding temperature is larger than its vaporisation temperature, $T_\mathrm{vap}$, then the pebbles do not reach the planet's surface for certain atmospheres and instead deposit all of their remaining spin at $r=R_\mathrm{vap}$. This assumes that frictional heat is negligible. For large objects friction is an important source of heat, but the small pebbles used in this study are slowed down much more effectively, meaning that they travel through the atmosphere at lower velocities and experience less friction \citep{Brouwers_2020}.

Evaporation only plays a role when the vaporisation radius $R_\mathrm{vap}$ is larger than the planet's radius. The vaporisation radius is given by \citep{Brouwers_2020}
\begin{equation}
        R_\mathrm{vap} = \frac{R'_\mathrm{B}}{(T_\mathrm{vap}/T_\mathrm{disc}-1)+R'_\mathrm{B}/R_\mathrm{rcb}},
\end{equation}
in which $R'_\mathrm{B}$ is the modified Bondi radius given by $R'_\mathrm{B}=\frac{\gamma-1}{\gamma}R_\mathrm{B}$ and $R_\mathrm{rcb}$ is the radiative-convective boundary, which is the boundary between the convective intermediate layer and the outer isothermal layer. Since the atmosphere in this study did not include an isothermal layer, we assumed that $R_\mathrm{rcb}=R_\mathrm{out}=\min(R_\mathrm{B},R_\mathrm{H})$. Assuming that the Bondi radius is smaller than the Hill radius, the vaporisation radius is given by
\begin{equation}
        R_\mathrm{vap} = \frac{G\mu(\gamma -1)}{k_\mathrm{b}(\gamma T_\mathrm{vap}-T_\mathrm{disc})}M_\mathrm{p}.
\end{equation}

The value of $T_\mathrm{vap}$ depends on the composition of the pebbles. Assuming a mainly silicate composition, an approximate temperature of 2500 K is appropriate. Given this temperature, pebble ablation only occurs in the atmospheres of massive planets. For a 0.1 Earth-mass planet $R_\mathrm{vap}\approx 0.4\ R_\mathrm{p}$, so ablation is not important. For the 0.5 Earth-mass planets in this study $R_\mathrm{vap}\approx 1.2\ R_\mathrm{p}$, meaning that ablation does occur. However, in terms of the Bondi radius, this vaporisation radius is at $0.022$ to $0.003\ R_\mathrm{B}$ at 1 and 50 au, respectively. As seen in Figs. \ref{fig:spin_afo_a} and \ref{fig:spin_Stk_vs_M}, in these situations the atmosphere is so thick that no angular momentum reaches the surface. Figure \ref{fig:stored_spin} shows that a significant portion of the spin of these pebbles is already stored in the inner 10\% of the atmosphere. Ablation will therefore not alter the results in this situation either. Since the Bondi radius and the vaporisation radius have the same proportionality to the planet's mass, there is no planet mass for which $R_\mathrm{vap}>0.1R_\mathrm{B}$. We can therefore confidently state that ablation will have a minimal influence on the results of this study.\\

The inclusion of the radiative outer layer, however, could influence the results more significantly for specific opacities. The pressure and density profiles in the radiative layer are exponential functions, meaning that the density in the outer regions increases more rapidly. The extent and influence of this layer depend on the assumed opacity. When assuming small opacities, such as a molecular opacity, the radiative zone is always important, unless the planet is both exceedingly small and close to the star \citep{Brouwers_2020}. However, pebble and dust accretion significantly increases the opacity of the atmosphere, leading to a dominance of the convective layer \citep{Brouwers_2021}. For accretion rates larger than about $10^{-6}$ M$_\Earth$/yr and planets smaller than about 2 M$_\Earth$, the atmosphere becomes fully convective \citep{Brouwers_2021}. Since the planets in this study did not exceed masses of 1 M$_\Earth$ and accretion rates between $10^{-6}$ and $10^{-5}$ M$_\Earth$/yr are typical \citep{Lambrechts_2014}, the radiative layer does not play a significant part.\\

The simulations in this study were performed in a planar (2D) configuration. In a 3D configuration, the spin asymmetry in shear-dominated PA is stronger and favoured more towards prograde spin. Especially for massive planets in a 3D setting, the asymmetry can become so large that planets are spun up well above their break-up spin \citep{Visser_2019}. The influence of the static atmosphere in a 3D model will likely be similar to the strong damping effect seen in the 2D case. In that way, the atmosphere might help prevent the planets from reaching spin rates that stop further growth. However, the atmospheric recycling discussed in the previous section only occurs in a 3D hydrodynamical model, in which it is essential that gas can stream in at different latitudes in the atmosphere. In 2D simulations, a stable atmosphere is formed with closed streamlines \citep{Ormel_2014_atm_rec_I,Ormel_2015_atm_rec_II}. To discuss the spin evolution in the atmosphere in Sect. \ref{sec:discussion-recycling}, we have mixed the results from 3D hydrodynamical simulations with the results from our 2D N-body simulations. To correctly quantify the spin of the planet and properly link the spin in the atmosphere to atmospheric recycling, the N-body simulations must be performed in 3D as well. However, since the qualitative trends of 2D and 3D simulations are similar \citep{Visser_2019}, we believe the comparison can still be made to get an idea of the resulting spin. In fact, the planar configuration, which effectively assumes that the pebbles have settled to a midplane, might be more appropriate for predicting the spins of the planets used in this study than a 3D configuration. The 3D configuration is applicable when the pebble scale height exceeds the impact radius for PA, which is only valid for small pebbles that experience strong turbulence and for small planets \citep{Visser_2019}.\\

\section{Conclusion}
\label{sec:conclusion}

We have analysed the influence of a static, adiabatic atmosphere on the spin build-up on a planet's surface due to PA in a global frame. The spin itself was determined by releasing 500 pebbles spread evenly throughout the collision cross-section and calculating the average angular momentum they deposit on the planet's surface. The key conclusions from this research are as follows.

\begin{enumerate}
        \item A static atmosphere dampens the spin frequency of a planet's surface. It does so by circularising the orbits of the pebbles around the planet, reducing the angular momentum they retain until the moment of impact with the planet. This is in contrast with the results of \citet{Takaoka_2023}, who analysed PA in the presence of a rapidly rotating atmosphere and find a strong prograde spin enhancement for all situations. In our extreme case, we find that nearly all individual spin contributions of the pebbles are forced to plateaus at $l_\mathrm{z}=\pm l_\mathrm{z,Kepler}=\pm\sqrt{GM_\mathrm{p}R_\mathrm{p}}$. For more massive planets and thicker atmospheres, there is a strong decay in the final stages of the orbit, leading to a less tangential impact and forcing the spin of the pebbles to sub-Keplerian plateaus. Generally, the atmosphere does not change the direction of the built-up spin.
        
        \item The spin reduction is seen for all distances to the star, Stokes numbers, and planet masses, provided the planet is massive enough that $R_\mathrm{B}>R_\mathrm{p}$. The more massive the planet, the more spin is absorbed in the atmosphere. For planets with a mass greater than 0.5 to 1.0 Earth mass, no angular momentum reaches the planet's surface, irrespective of the orbital radius of the planet or the Stokes number of the pebbles. This means that an Earth-like or super-Earth-like planet with a static atmosphere cannot have been spun up by PA, at least not in the final stages of its growth, unless significant amounts of spin deposited in the atmosphere were transported to the surface.
        
        \item The amount of spin stored in the inner 10\% of the atmosphere is equal to or greater than the difference in spin between the atmosphere and the atmosphere-less planet surfaces for any distance to the star for a planet of 0.1 Earth masses.
        
        \item The damping of the atmosphere is slightly asymmetrical, with retrograde orbiting pebbles losing on average 1\% more angular momentum to the atmosphere than prograde orbiting pebbles. In situations in which the prograde and retrograde spin contributions of PA fully balance out and the mean spin without an atmosphere is zero, this leads to an absolute spin increase due to an atmosphere. This increase can be up to $0.06-0.12\  \omega_\mathrm{crit}$ for large Stokes numbers around a small planet.
        
        \item Three-dimensional simulations without an atmosphere predict spin values for certain asteroids and planets of twice their break-up spin \citep{Visser_2019}. Assuming that the damping effect of the atmosphere can be generalised to a 3D situation, an atmosphere might be necessary to prevent excessive spin build-up and allow for further growth.
\end{enumerate}
This study shows that an atmosphere plays a significant part in spin transfer during PA, especially in more massive planets. Because of the difference between our results and those of \citet{Takaoka_2023}, our study also shows that assumptions about the rotation rate of the atmosphere play a vital role in the final spin state of the planet. In the extreme case of the rapidly rotating atmospheres of \citet{Takaoka_2023}, massive planets build up prograde spin close to their break-up spin, while in our extreme case of static atmospheres, those same planets build up no spin.

\section*{Acknowledgements}
We would like to thank Marc Brouwers for the valuable discussions about the atmospheric models and for his feedback on the manuscript. We would also like to thank the anonymous referee for their commentary, which significantly improved the clarity of this paper. R.G. Visser acknowledges funding from the Dutch Research Council (NWO project no. ALWGO/15-01).

\bibliographystyle{aa}
\bibliography{references}

\begin{thebibliography}{51}
\expandafter\ifx\csname natexlab\endcsname\relax\def\natexlab#1{#1}\fi

\bibitem[{{Batchelor}(1967)}]{Batchelor_1967}
{Batchelor}, G.~W. 1967, An introduction to fluid dynamics (Cambridge
  University Press)

\bibitem[{Benvenuto \& Brunini(2005)}]{Benvenuto2005}
Benvenuto, O.~G. \& Brunini, A. 2005, \mnras, 356, 1383

\bibitem[{Bottke {et~al.}(2005)Bottke, Durda, Nesvorný, Jedicke, Morbidelli,
  Vokrouhlický, \& Levison}]{Bottke_2005}
Bottke, W.~F., Durda, D.~D., Nesvorný, D., {et~al.} 2005, \icarus, 175, 111

\bibitem[{Brouwers \& Ormel(2020)}]{Brouwers_2020}
Brouwers, M.~G. \& Ormel, C.~W. 2020, \aap, 634, A15

\bibitem[{{Brouwers} {et~al.}(2021){Brouwers}, {Ormel}, {Bonsor}, \&
  {Vazan}}]{Brouwers_2021}
{Brouwers}, M.~G., {Ormel}, C.~W., {Bonsor}, A., \& {Vazan}, A. 2021, \aap,
  653, A103

\bibitem[{{Brouwers} {et~al.}(2018){Brouwers}, {Vazan}, \&
  {Ormel}}]{Brouwers_2018}
{Brouwers}, M.~G., {Vazan}, A., \& {Ormel}, C.~W. 2018, \aap, 611, A65

\bibitem[{Cimerman {et~al.}(2017)Cimerman, Kuiper, \& Ormel}]{Cimerman_2017}
Cimerman, N.~P., Kuiper, R., \& Ormel, C.~W. 2017, \mnras, 471, 4662

\bibitem[{D'Angelo \& Bodenheimer(2013)}]{D'Angelo_2013}
D'Angelo, G. \& Bodenheimer, P. 2013, \apj, 778, 77

\bibitem[{Dobrovolskis(1980)}]{DOBROVOLSKIS_1980}
Dobrovolskis, A.~R. 1980, \icarus, 41, 18

\bibitem[{Dones \& Tremaine(1993a)}]{Dones_1993a}
Dones, L. \& Tremaine, S. 1993a, \icarus, 103, 67

\bibitem[{Dones \& Tremaine(1993b)}]{Dones_1993b}
Dones, L. \& Tremaine, S. 1993b, Science, 259, 350

\bibitem[{{Hayashi} {et~al.}(1985){Hayashi}, {Nakazawa}, \&
  {Nakagawa}}]{Hayashi}
{Hayashi}, C., {Nakazawa}, K., \& {Nakagawa}, Y. 1985, in Protostars and
  Planets II, ed. D.~C. {Black} \& M.~S. {Matthews}, 1100--1153

\bibitem[{Hubickyj {et~al.}(2005)Hubickyj, Bodenheimer, \&
  Lissauer}]{HUBICKYJ2005415}
Hubickyj, O., Bodenheimer, P., \& Lissauer, J.~J. 2005, \icarus, 179, 415

\bibitem[{Ida \& Nakazawa(1990)}]{Ida_1990}
Ida, S. \& Nakazawa, K. 1990, \icarus, 86, 561

\bibitem[{Johansen \& Lacerda(2010)}]{Johansen_Lacerda_2010}
Johansen, A. \& Lacerda, P. 2010, \mnras, 404, 475

\bibitem[{{Johansen} {et~al.}(2015){Johansen}, {Mac Low}, {Lacerda}, \&
  {Bizzarro}}]{Johansen_2015b}
{Johansen}, A., {Mac Low}, M.-M., {Lacerda}, P., \& {Bizzarro}, M. 2015,
  Science Advances, 1, 1500109

\bibitem[{Kokubo \& Ida(2007)}]{Kokubo_2007}
Kokubo, E. \& Ida, S. 2007, \apj, 671, 2082

\bibitem[{Kurokawa \& Tanigawa(2018)}]{Kurokawa_2018}
Kurokawa, H. \& Tanigawa, T. 2018, \mnras, 479, 635

\bibitem[{{Lambrechts} \& {Johansen}(2012)}]{Lambrechts_2010}
{Lambrechts}, M. \& {Johansen}, A. 2012, \aap, 544, A32

\bibitem[{Lambrechts \& Johansen(2014)}]{Lambrechts_2014}
Lambrechts, M. \& Johansen, A. 2014, \aap, 572, A107

\bibitem[{{Laskar} \& {Robutel}(1993)}]{Laskar_1993}
{Laskar}, J. \& {Robutel}, P. 1993, \nat, 361, 608

\bibitem[{Lee \& Chiang(2015)}]{Lee_2015}
Lee, E.~J. \& Chiang, E. 2015, \apj, 811, 41

\bibitem[{Lissauer {et~al.}(1997)Lissauer, Berman, Greenzweig, \&
  Kary}]{Lissauer_1997}
Lissauer, J.~J., Berman, A.~F., Greenzweig, Y., \& Kary, D.~M. 1997, \icarus,
  127, 65

\bibitem[{Lissauer \& Kary(1991)}]{Lissauer_1991}
Lissauer, J.~J. \& Kary, D.~M. 1991, \icarus, 94, 126

\bibitem[{Liu \& Ormel(2018)}]{Liu&Ormel2018}
Liu, B. \& Ormel, C.~W. 2018, \aap, 615, A138

\bibitem[{Machida {et~al.}(2008)Machida, Kokubo, ichiro Inutsuka, \&
  Matsumoto}]{Machida_2008}
Machida, M.~N., Kokubo, E., ichiro Inutsuka, S., \& Matsumoto, T. 2008, \apj,
  685, 1220

\bibitem[{Miguel \& Brunini(2010)}]{Miguel_2010}
Miguel, Y. \& Brunini, A. 2010, \mnras, 406, 1935

\bibitem[{{Morbidelli} {et~al.}(2015){Morbidelli}, {Lambrechts}, {Jacobson}, \&
  {Bitsch}}]{Morbidelli_2015}
{Morbidelli}, A., {Lambrechts}, M., {Jacobson}, S., \& {Bitsch}, B. 2015,
  \icarus, 258, 418

\bibitem[{Nakagawa {et~al.}(1986)Nakagawa, Sekiya, \& Hayashi}]{NAKAGAWA1986}
Nakagawa, Y., Sekiya, M., \& Hayashi, C. 1986, \icarus, 67, 375

\bibitem[{{Okamura} \& {Kobayashi}(2021)}]{okamura2021}
{Okamura}, T. \& {Kobayashi}, H. 2021, \apj, 916, 109

\bibitem[{{Ormel}(2017)}]{Ormel_2017_PA}
{Ormel}, C.~W. 2017, {The Emerging Paradigm of Pebble Accretion}, ed.
  M.~{Pessah} \& O.~{Gressel}, Vol. 445, 197

\bibitem[{{Ormel} \& {Klahr}(2010)}]{Ormel_2010}
{Ormel}, C.~W. \& {Klahr}, H.~H. 2010, \aap, 520, A43

\bibitem[{Ormel {et~al.}(2014)Ormel, Kuiper, \& Shi}]{Ormel_2014_atm_rec_I}
Ormel, C.~W., Kuiper, R., \& Shi, J.-M. 2014, \mnras, 446, 1026

\bibitem[{Ormel {et~al.}(2015{\natexlab{a}})Ormel, Shi, \&
  Kuiper}]{Ormel_2015_atm_rec_II}
Ormel, C.~W., Shi, J.-M., \& Kuiper, R. 2015{\natexlab{a}}, \mnras, 447, 3512

\bibitem[{Ormel {et~al.}(2015{\natexlab{b}})Ormel, Shi, \&
  Kuiper}]{Ormel_2015b}
Ormel, C.~W., Shi, J.-M., \& Kuiper, R. 2015{\natexlab{b}}, \mnras, 447,
  3512–3525

\bibitem[{Pollack {et~al.}(1996)Pollack, Hubickyj, Bodenheimer, Lissauer,
  Podolak, \& Greenzweig}]{Pollack_1996}
Pollack, J.~B., Hubickyj, O., Bodenheimer, P., {et~al.} 1996, \icarus, 124, 62

\bibitem[{Popovas {et~al.}(2018)Popovas, Nordlund, Ramsey, \&
  Ormel}]{Popovas_2018}
Popovas, A., Nordlund, {\AA}., Ramsey, J.~P., \& Ormel, C.~W. 2018, \mnras,
  479, 5136

\bibitem[{{Rein} \& {Liu}(2012)}]{rebound}
{Rein}, H. \& {Liu}, S.~F. 2012, \aap, 537, A128

\bibitem[{{Rein} \& {Spiegel}(2015)}]{reboundias15}
{Rein}, H. \& {Spiegel}, D.~S. 2015, \mnras, 446, 1424

\bibitem[{{Ricci} {et~al.}(2010){Ricci}, {Testi}, {Natta}, \&
  {Brooks}}]{Ricci_2010}
{Ricci}, L., {Testi}, L., {Natta}, A., \& {Brooks}, K.~J. 2010, \aap, 521, A66

\bibitem[{Steinberg \& Sari(2015)}]{Steinberg_2015}
Steinberg, E. \& Sari, R. 2015, \aj, 149, 124

\bibitem[{{Takaoka} {et~al.}(2023){Takaoka}, {Kuwahara}, {Ida}, \&
  {Kurokawa}}]{Takaoka_2023}
{Takaoka}, K., {Kuwahara}, A., {Ida}, S., \& {Kurokawa}, H. 2023, arXiv
  e-prints, arXiv:2303.15098

\bibitem[{{Tamayo} {et~al.}(2020){Tamayo}, {Rein}, {Shi}, \&
  {Hernandez}}]{reboundx}
{Tamayo}, D., {Rein}, H., {Shi}, P., \& {Hernandez}, D.~M. 2020, \mnras, 491,
  2885

\bibitem[{{Testi} {et~al.}(2003){Testi}, {Natta}, {Shepherd}, \&
  {Wilner}}]{Testi2003}
{Testi}, L., {Natta}, A., {Shepherd}, D.~S., \& {Wilner}, D.~J. 2003, \aap,
  403, 323

\bibitem[{{Visser} \& {Brouwers}(2022)}]{Visser_2022}
{Visser}, R.~G. \& {Brouwers}, M.~G. 2022, \aap, 663, A164

\bibitem[{Visser \& Ormel(2016)}]{Visser_2016}
Visser, R.~G. \& Ormel, C.~W. 2016, \aap, 586, A66

\bibitem[{Visser {et~al.}(2019)Visser, Ormel, Dominik, \& Ida}]{Visser_2019}
Visser, R.~G., Ormel, C.~W., Dominik, C., \& Ida, S. 2019, \icarus, 335

\bibitem[{{Weidenschilling}(1977{\natexlab{a}})}]{Weidenschilling_1977a}
{Weidenschilling}, S.~J. 1977{\natexlab{a}}, \mnras, 180, 57

\bibitem[{{Weidenschilling}(1977{\natexlab{b}})}]{Weidenschilling_1977b}
{Weidenschilling}, S.~J. 1977{\natexlab{b}}, \apss, 51, 153

\bibitem[{Wetherill(1985)}]{Wetherill_1958}
Wetherill, G.~W. 1985, Science, 228, 877

\bibitem[{Wilner {et~al.}(2005)Wilner, D'Alessio, Calvet, Claussen, \&
  Hartmann}]{Wilner_2005}
Wilner, D.~J., D'Alessio, P., Calvet, N., Claussen, M.~J., \& Hartmann, L.
  2005, \apj, 626, L109

\end{thebibliography}

\begin{appendix}
\section{Numerical model and initial conditions}
\label{sec:appendix_A}

The equations of motion of the pebbles in this study were integrated using the REBOUND N-body code \citep{rebound} with the additional physics packages from REBOUNDX \citep{reboundx}. The integrator was a 15$^\mathrm{th}$ order Gauss-Radau integrator with adaptive step-size control (AIS15) \citep{reboundias15} with a tolerance of $10^{-6}$.\\ 

As mentioned in Sect. \ref{sec:EoM}, the pebbles were initialised along a circular orbit exterior to the planet's orbit. More precisely, the pebbles were released from within the collisional cross-section, which is the sub-domain of the aforementioned orbit from within which the pebbles eventually collide with the planet (see Fig.\ref{fig:simsetup}). The initial radial and azimuthal velocities were given by \citep{Liu&Ormel2018}
\begin{equation}
        \begin{cases}
                v_{r,0}=v_{r,\infty}=-\cfrac{2 v_\mathrm{hw} \tau_\mathrm{s}}{1+\tau_\mathrm{s}^2},\\
                v_{\phi,0}=v_{\phi,\infty}=V_\mathrm{k} - \cfrac{v_\mathrm{hw}}{1+\tau_\mathrm{s}^2},
        \end{cases}
\end{equation}
in which $\tau_\mathrm{s}$ is the Stokes number of the pebbles, $v_\mathrm{hw}$ is the headwind velocity and $V_\mathrm{k}$ is the velocity of a circular Keplerian orbit.

\begin{figure}
        \centering
        \includegraphics[width=\linewidth]{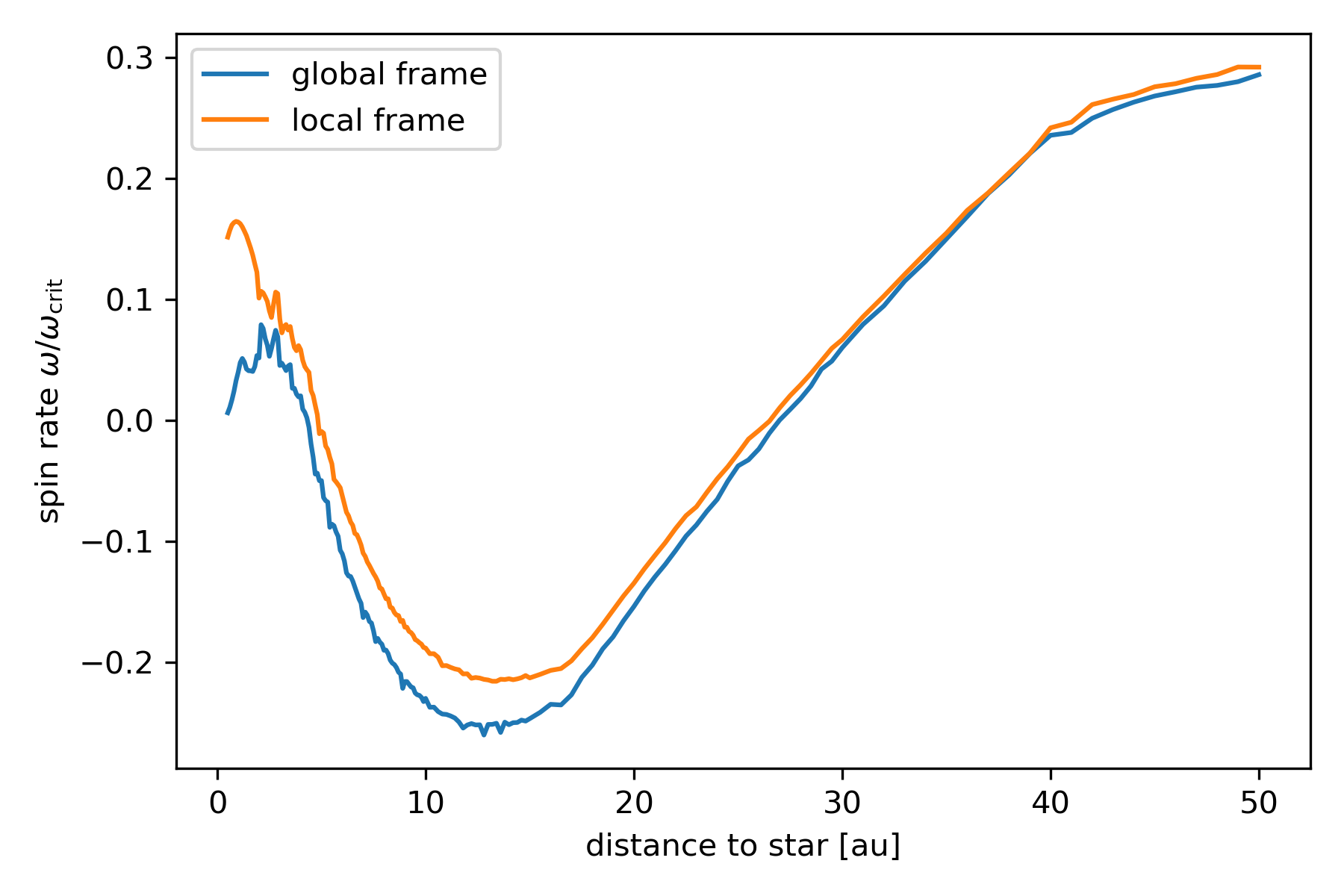}
        \caption{Comparison between the spin frequency in the global (heliocentric) and local (shearing sheet) frame for an atmosphere-less 0.1 Earth-mass planet and pebbles with a Stokes number of 0.1. Especially in the inner disc ($<20$ au), the difference between the global and local frame is significant.}
        \label{fig:HC_vs_SS}
\end{figure}

As for the initial position of the pebbles, the pebbles needed to be released far enough from the planet that the planetary gravity at the moment of release was negligible. This is the reason why the simulations in this study were performed in a global frame instead of in a local frame. Because of the large planet masses required to form atmospheres, the shearing sheet approximation used in the local frame was no longer valid. In a shearing sheet approximation, the release distances $x_0$ and $y_0$ should be much less than the orbital distance to the central body $r_0$ \citep{Liu&Ormel2018}. $y_0$ is given by \citep{Visser_2019, Ormel_2010}
\begin{equation}
        y_0=100\sqrt{GM_\mathrm{p}t_\mathrm{s}/v_\mathrm{hw}},
\end{equation}
in which $M_\mathrm{p}$ is the planet's mass and $t_\mathrm{s}$ is the stopping time of the pebbles. For the pebbles to be released in a gas drag-governed condition, unperturbed by the planetary gravity, $y_0$ had to be of the order of multiple astronomical units for planets of 0.1 Earth masses. For these scales, the curvature of the disc starts to significantly influence the results and the shearing sheet approximation breaks down, as can be seen in Fig.\ref{fig:HC_vs_SS}.\\

Our simulations were therefore performed in a global frame. The planet was initiated on a circular orbit with an orbital radius a$_\mathrm{plan}$. To ensure the unperturbed initial state requirement was met, the pebbles were released at a minimum distance from the planet of 5 Hill radii ($R_\mathrm{H}$). Because of the symmetry of the circular planetary orbit, the pebbles could then be released along a ring in the ecliptic plane with a radius $r_0=\mathrm{a}_\mathrm{plan} +5R_\mathrm{H}$, varying only in initial azimuthal coordinate $\phi_0$
\footnote{This only holds for circular planetary orbits. In this system, the amount of spin a pebble will transfer to the planet is fully determined by the azimuthal separation $\Delta \phi_\mathrm{pebb,plan}$ of the pebble to the planet at some arbitrary point of measurement. A pebble released at $r>r_0$ will transfer the same amount of angular momentum to the planet as a pebble released at $r=r_0$ with the same angular separation at $r_0$. Even though the phase of the planet at the moment of impact will be different, the amount of transferred spin will be the same. This is not valid for planets with elliptical orbits, since the orientation of the ellipse and the phase of the planet at the moment of impact do influence the amount of spin transferred. In that case, a collisional cross-section must be calculated as a function of both $\phi$ and $r$.}
\citep{Liu&Ormel2018}. The advantage of this approach is that it reduced the collisional cross-section to one dimension, specifically to a sub-domain of $\phi\in[0,2\pi)$, independent of $r$. This allowed a bifurcation algorithm to be used to find the collisional cross-section with the desired accuracy, $10^{-5}$ rad.\\

However, the downside of the global approach compared to a local approach is the numerical error. There is a large difference in scale in the heliocentric simulations. The scale of full planetary orbits of astronomical units is combined with the scale of close encounters between pebbles and the planet's surface. This means that for a planet at 10 au and a pebble close to the surface, of the 15 to 17 significant digits allowed by the machine precision, the first 5 significant digits are taken up by the scale of the planet's orbit and only the final 10 can be used for the position of the pebble with respect to the planet. This increases the integration error, which leads to an exceedingly small step size in the AIS15 integrator, which in turn leads to unreasonably long integration times. To prevent such problems, the error tolerance of AIS15 was increased from its default value of $10^{-9}$ to $10^{-6}$. However, this increases the uncertainty of the results. It is therefore advisable for future studies to create a hybrid model that starts in a global frame and transforms to a local frame co-rotating with the planet as soon as the pebbles enter the Hill sphere, as was done by \citet{Liu&Ormel2018}.\\  

On a separate and final note, at the start of the simulations, the pebble state vector was transformed from polar to Cartesian coordinates and the system was centred on the centre of mass (COM). The latter was important since, in REBOUND simulations, the planet and the central star are both physical particles with mutual gravitational interactions, meaning that they both rotate around the centre of mass, rather than that the planet revolves around the star. The rotation of the non-inertial heliocentric frame in REBOUND had a measurable effect on the spin transfer, when not corrected for.\\   

\end{appendix}

\end{document}